\documentclass[12pt]{article}

\usepackage{graphics}
\usepackage{amsmath}
\usepackage{epsfig}
\usepackage{cite}
 
\title{Inclusive Higgs boson production at the LHC in the $k_T$-factorization approach}

\author{N.A.~Abdulov$^{1}$, A.V.~Lipatov$^{2,\,3}$, M.A.~Malyshev$^2$}

\begin{document}

\maketitle

\begin{center}

{\it $^1$Faculty of Physics, Lomonosov Moscow State University, 119991 Moscow, Russia}\\
{\it $^2$Skobeltsyn Institute of Nuclear Physics, Lomonosov Moscow State University, 119991 Moscow, Russia}\\
{\it $^3$Joint Institute for Nuclear Research, Dubna 141980, Moscow Region, Russia}

\end{center}

\vspace{0.5cm}

\begin{center}

{\bf Abstract }

\end{center} 

\indent
We investigate the inclusive Higgs boson production in proton-proton collisions 
at the CERN LHC conditions using the $k_T$-factorization approach. 
Our analysis is based on the dominant off-shell gluon-gluon fusion subprocess (where 
the transverse momenta of initial gluons are taken into account)
and covers $H \to \gamma \gamma$, $H \to ZZ^* \to 4l$ (where $l = e,\mu$) and 
$H \to W^+ W^- \to e^\pm \mu^\mp \nu \bar \nu$ decay channels.
The transverse momentum dependent (or unintegrated) gluon densities in a proton were derived from
Ciafaloni-Catani-Fiorani-Marchesini equation or, alternatively, were
chosen in accordance with Kimber-Martin-Ryskin prescription.
We estimate the theoretical uncertainties of our
calculations and compare our results with next-to-next-to-leading-order plus 
next-to-next-to-leading-logarithmic ones obtained 
using collinear QCD factorization.
Our predictions agree well with the latest experimental data taken 
by the CMS and ATLAS Collaborations at $\sqrt s = 8$ and $13$~TeV.

\vspace{1.0cm}

\noindent
PACS number(s): 12.38.Bx, 14.80.Bn

\newpage

\section{Introduction} \indent

With the startup of the Large Hadron Collider (LHC), high energy physics entered a
new era. A great triumph of the Standard Model (SM) is the discovery of the 
Higgs boson in $2012$\cite{1,2}. 
The Higgs boson $H$ was predicted more than $50$ years ago as a consequence of the 
electroweak symmetry breaking mechanism in the SM. 
This mechanism introduces a single 
complex scalar field doublet, which gives masses to $W$ and $Z$ bosons
and to fundamental fermions through Yukawa interaction\cite{3,4,5}.
The SM Higgs boson is the physical neutral
scalar field which is the only remaining part of this doublet
after spontaneous symmetry breaking.
In extensions of SM there are 
additional charged and neutral scalar or pseudoscalar Higgs particles.
Theoretical and experimental investigations of the Higgs boson
production cross sections and its decay rates are an important test 
for possible deviations from the SM expectations\cite{6,7,8,9,10}.

Recently the CMS and ATLAS Collaborations have reported their measurements\cite{11,12,13,14,15,16}
of the inclusive Higgs boson total and differential cross sections
at $\sqrt s = 8$~TeV in 
the $H \to \gamma \gamma$, $H \to ZZ^* \to 4l$ (with $l = e,\mu$) 
and $H \to W^+W^- \to e^\pm \mu^\mp \nu \bar \nu$ decay channels.
Moreover, preliminary data collected at $\sqrt s = 13$~TeV have become available\cite{17,18,19,20}. 
The measured observables, such as distributions on the 
transverse momentum, rapidity or scattering angle of decay 
particles, allow to probe fundamental properties 
of the Higgs boson (for example, spin and couplings to gauge bosons and fermions)
and can be used to investigate the gluon dynamics in a proton
since the dominant mechanism of inclusive Higgs production at the LHC is the
gluon-gluon fusion\footnote{The gluon-gluon fusion and 
weak boson fusion (namely, $qq \to qqH$ subprocess via t-channel exchange 
of a $W$ or $Z$ bosons) are also expected to be the dominant sources
of semi-inclusive Higgs production at the LHC.}\cite{6,7,8,9,10}.
Corresponding total and differential cross sections measured at $\sqrt s = 8$~TeV are higher than the SM estimations,
obtained at next-to-next-to-leading order (NNLO)\cite{21,22,23,24,25,26} and matched 
with soft-gluon resummation carried out up to next-to-next-to-leading logarithmic 
accuracy (NNLL)\cite{27,28}, although no significant deviations from the perturbative 
Quantum Chromodynamics (pQCD) predictions\footnote{The next-to-leading order perturbative 
electroweak corrections to the Higgs production cross section are available\cite{29,30,31,32,33}.} 
within the experimental and theoretical uncertainties are observed\cite{11,12,13,14,15,16}.
The same conclusion was made\cite{17,18,19,20} for preliminary data taken 
by the CMS and ATLAS Collaborations at $\sqrt s = 13$~TeV.
The latter were compared with the \textsc{nnlops} calculations\cite{34,35} normalized to 
N$^3$LO predictions\cite{36,37,38} for gluon-gluon fusion subprocess. The \textsc{nnlops} tool
provides parton-level events at NNLO accuracy and is interfaced
to the \textsc{pythia8} event generator\cite{39} for parton showering, hadronization and 
multiple parton interactions.

In the present study we give a systematic QCD analysis of the latest CMS\cite{11,12,13,17,18} and 
ATLAS\cite{14,15,16,19,20} data on the inclusive Higgs production
in diphoton, four-lepton and $H \to W^+W^- \to e^\pm \mu^\mp \nu \bar \nu$
decay modes collected at $\sqrt s = 8$ and $13$~TeV using the 
$k_T$-factorization approach\cite{40,41}.
The $k_T$-factorization approach is based on the Balitsky-Fadin-Kuraev-Lipatov (BFKL)\cite{42} or 
Ciafaloni-Catani-Fiorani-Marchesini (CCFM)\cite{43} gluon
evolution equations, which resum large logarithmic terms 
proportional to $\ln s \sim \ln 1/x$, important at high energies (or,
equivalently, at small proton longitudinal momentum fraction $x$ carried by gluons).
The CCFM equation takes into account additional terms proportional to $\ln 1/(1 - x)$
and is almost equivalent to the BFKL equation in the
limit of asymptotic energies, but also similar to the 
conventional Dokshitzer-Gribov-Lipatov-Altarelli-Parisi (DGLAP)\cite{44} 
scenario for large $x$ and high scale $\mu^2$.
For inclusive Higgs production at the LHC, typical $x$
values are $x \sim m_H/\sqrt s \sim 0.008 - 0.015$ (for Higgs mass $m_H \sim 125$~GeV), 
so that one can reach the low $x$ domain where
the BFKL-like evolution is expected to be valid.
Additionally, we see certain advantages in the fact that, 
even with the leading-order (LO) partonic amplitudes, 
a large piece of higher order corrections
(namely, part of NLO + NNLO + ... terms corresponding to real gluon
emissions in initial state)
are included by using transverse momentum dependent (TMD) gluon 
densities. Besides that, the latter absorb the effects of soft gluon resummation\cite{45}, that
regularises the infrared divergences and makes our predictions valid even
at low transverse momenta.
More detailed descriptions of the $k_T$-factorization
formalism can be found, for example, in reviews\cite{46}.

The $k_T$-factorization approach has been already applied to the 
inclusive Higgs boson production\cite{45,47,48,49,50,51,52,53,54}.
So, the effective Lagrangian\cite{55,56} for the Higgs coupling 
to gluons (valid in the large top quark mass limit, $m_t \to \infty$) 
was used\cite{45,47,49,50,51,52,53,54} to calculate the amplitude of dominant gluon-gluon fusion 
subprocess, whereas finite top mass $m_t$ effects in the triangle quark loop 
were investigated\cite{48}.
The Kimber-Martin-Ryskin (KMR)\cite{57} 
prescription for the TMD gluon density in a proton 
(where the gluon transverse momentum is generated 
at the last evolution step) was applied\cite{49} and the
simplified solution of the CCFM equation in the single loop approximation (where the 
small-$x$ effects are neglected) was used\cite{45}.
In the framework of Monte-Carlo generator \textsc{cascade}\cite{58}
the off-shell production amplitude\cite{50} was used with the full CCFM evolution\cite{51}.
Recently, it was demostrated\cite{52} that the $k_T$-factorization approach supplemented with the CCFM gluon 
dynamics is able to describe first (preliminary) data\cite{59} on the inclusive 
Higgs production in the diphoton decay mode\footnote{The preliminary ATLAS data\cite{59} 
on the Higgs boson transverse momentum distribution were discussed also\cite{54}. 
However, the calculations\cite{54} are based on rather old CCFM-evolved TMD gluon density function
and, in our opinion, suffer from double counting.} taken by the ATLAS 
Collaboration at the LHC. 
The effect of taking into account higher-order corrections
in the $k_T$-factorization approach at LO was pointed out\cite{47,49,52,53}.
The CMS\cite{12} and ATLAS data\cite{14} for Higgs boson production
in the four-lepton decay mode were considered\cite{53}. 

Our present consideration is based on the off-shell amplitude of the gluon-gluon fusion 
subprocess $g^* g^* \to H$\cite{47}. The latter was extended further
to the subsequent diphoton\cite{52} and four-lepton Higgs boson decays\cite{53}.
Below we will derive the expressions for off-shell 
$g^* g^* \to H \to W^+W^- \to e^\pm \mu^\mp \nu \bar \nu$
and $g^* g^* \to H \to ZZ^* \to 4l$ (where $l = e, \mu$)
amplitudes (independently from\cite{53}).
Then, to calculate the Higgs boson production cross section we convolute these amplitudes
with the TMD gluon densities in a proton, taken from the
numerical solution of the CCFM equation\cite{60}.
As an alternative choice, we will use the TMD gluon densities evaluated 
in accordance with the KMR prescription\cite{57}.
Our main motivation is that the latest
CMS\cite{11,13} and ATLAS\cite{14,16} data taken at $\sqrt s = 8$~TeV (referring 
to $H\to \gamma \gamma$ and $H \to W^+W^- \to e^\pm \mu^\mp \nu \bar \nu$ 
decay channels) as well as preliminary data\cite{17,18,19,20} obtained at $\sqrt s = 13$~TeV
have not been analysed yet in the framework of $k_T$-factorization.
Additionally, detailed studying of the Higgs transverse momentum distributions
in the different kinematical regimes of different decay channels could impose 
constraints on the TMD gluon density (see\cite{47,52,53}).

The outline of our paper is following. In Section~2 we recall 
the basic formulas of $k_T$-factorization approach and briefly describe 
the calculation steps. In Section 3 we present our
numerical results and discussion. Section~4 contains our conclusions.

\section{The model} \indent

Let us start from a short review of the calculation steps.
We describe first the evaluation of $g^* g^* \to H \to ZZ^* \to 4l$ and 
$g^*g^* \to H \to W^+W^- \to e^\pm \mu^\mp \nu \bar \nu$ off-shell 
production amplitudes.
The effective Lagrangian for the Higgs boson 
coupling to gluons in the limit of large top quark mass 
$m_t \to \infty$ reads\cite{55,56}
\begin{equation}
  {\cal L}_{ggH} = {\alpha_s\over 12\pi} \left(G_F \sqrt 2\right)^{1/2} G_{\mu \nu}^a G^{a\, \mu \nu} H,
\end{equation}

\noindent 
where $G_F$ is the Fermi coupling constant, $G_{\mu \nu}^a$ is the gluon field strength tensor and
$H$ is the Higgs scalar field. The large $m_t$ approximation is valid 
to an accuracy of few percents in the mass range $m_H < 2 m_t$, and,
of course, is applicable at the $m_H \sim 125$~GeV\cite{11,12,13,14,15,16,17,18,19,20}.
The triangle vertex for two off-shell gluons having four-momenta
$k_1$ and $k_2$ and color indices $a$ and $b$ thus takes the form\cite{55,56}:
\begin{equation}
  T^{\mu \nu,\, ab}_{ggH}(k_1,k_2) = i \delta^{ab} {\alpha_s\over 3\pi} \left(G_F \sqrt 2\right)^{1/2} \left[k_2^\mu k_1^\nu - (k_1 \cdot k_2) g^{\mu \nu}\right].
\end{equation}

\noindent 
Using~(2) and taking into account the non-zero transverse momenta of initial gluons
$k_1^2 = - {\mathbf k}_{1T}^2 \neq 0$ and $k_2^2 = - {\mathbf k}_{2T}^2 \neq 0$, 
one can easily obtain the off-shell production amplitudes squared for considered subprocesses.
The latter can be written in a compact form:
\begin{equation}
  \displaystyle |\mathcal{\bar M}|^2 = {8\over 9} {\alpha_{s}^2\over \pi^2} \, G_F \sqrt{2} \, (4\pi\alpha)^3 m_Z^2 \, C_V { (\hat s + {\mathbf p}_T^2)^2 \over (\hat s - m_H^2)^2 + m_H^2 \Gamma_H^2} \cos^2\phi \, \times \atop {
  \displaystyle \times \, {2 g_{(V)L}^2 \, g_{(V)R}^{2} \, (p_1 \cdot p_4)(p_2 \cdot p_3) + ( g_{(V)L}^4 + g_{(V)R}^4 ) (p_1 \cdot p_3)(p_2 \cdot p_4) \over [(q_1^2 - m_V^2)^2 + \Gamma_V^2 m_V^2] [(q_2^2 - m_V^2)^2 + \Gamma_V^2 m_V^2]} },
\end{equation}

\noindent 
where we have neglected the masses of final-state leptons\footnote{We do not consider 
here the case of identical leptons in the final state and 
calculate its contribution in the same manner as for distinct leptons.
This assumption is based on experimental kinematics cuts, which 
almost eliminate interference effects (see discussion in Section~3).}. 
The symbol $V$ denotes $Z$ or $W$ bosons, $p_1$ and $p_3$ are their decay leptons four-momenta, $p_2$ and $p_4$ are 
the antileptons four-momenta, $\hat s = (k_1 + k_2)^2$, the transverse momentum of
the Higgs particle is ${\mathbf p}_T = {\mathbf k}_1 + {\mathbf k}_2$, $\Gamma_H$ is its
full decay width, $\phi$ is the azimuthal angle between the transverse
momenta of initial gluons, $q_1^2$ and $q_2^2$ are the virtualities of the intermediate $Z$ or $W$ bosons,
$m_Z$, $m_W$, $\Gamma_Z$ and $\Gamma_W$ are their masses and full decay widths, respectively.
The constants $C_Z$ and $C_W$ are given by
\begin{equation}
  C_Z = {4\over \sin^6 2\theta_W},
\end{equation}
\begin{equation}
  C_W = {\cot^2 \theta_W \over 64 \sin^4 2\theta_W},
\end{equation}

\noindent 
where $\theta_W$ is the Weinberg mixing angle. The left and right weak current constants read:
\begin{equation}
  g_{(Z)L} = - {1\over 2} + \sin^2 \theta_W, \quad g_{(W)R} = 1,
\end{equation}
\begin{equation}
  g_{(Z)R} = \sin^2 \theta_W, \quad g_{(W)L} = 0.
\end{equation}

\noindent
The propagators of the intermediate Higgs and electroweak bosons are taken in the Breit-Wigner form to 
avoid any artificial singularities in the numerical calculations. 
According to the $k_T$-factorization prescription\cite{40,41}, the summation over the 
polarizations of initial off-shell gluons is carried out with
\begin{equation}
  \sum \epsilon^\mu \epsilon^{* \nu} = { {\mathbf k}_T^\mu {\mathbf k}_T^\nu \over {\mathbf k}_T^2}.
\end{equation}

\noindent
In the limit ${\mathbf k}_T \to 0$ this expression converges to the ordinary
one after averaging on the azimuthal angle. In all other respects
the calculations are quite straightforward and follow the standard QCD Feynman rules.
In the case of Higgs four-lepton decay $H \to ZZ^* \to 4l$, the obtained expression~(3)
coincides with the one\cite{53}.
The off-shell production amplitude for $g^* g^* \to H \to \gamma \gamma$ 
subprocess was calculated earlier\cite{52}.

To calculate the cross sections of the considered processes
in the $k_T$-factorization approach one should convolute corresponding 
off-shell partonic cross sections
with the TMD gluon densities in a proton. 
Our master formula for $H \to ZZ^* \to 4l$ and $H \to W^+W^- \to e^\pm \mu^\mp \nu \bar \nu$ decay 
channels reads:
\begin{equation}
  \displaystyle \sigma = {1\over (2\pi)^8} \int {\lambda^{1/2}(\hat s, q_1^2, q_2^2) \over 512 \, x_1 x_2 s \hat s \, \lambda^{1/2}(\hat s, k_1^2, k_2^2) } \, f_g(x_1,{\mathbf k}_{1T}^2,\mu^2) f_g(x_2,{\mathbf k}_{2T}^2,\mu^2) |{\cal \bar M}^2| \, \times \atop {
  \displaystyle \times \, d{\mathbf k}_{1T}^2 d{\mathbf k}_{2T}^2 dy dq_1^2 dq_2^2 d\hat s d\Omega^* d\Omega_1^* d\Omega_2^* {d\phi_1 \over 2\pi} {d\phi_2 \over 2\pi}},
\end{equation}

\noindent
where $f_g(x,{\mathbf k}_{T}^2,\mu^2)$ is the TMD gluon density, $s$ is the total center-of-mass energy,
$y$ is the Higgs boson rapidity,
$\Omega^*$ is the decay solid angle of a vector boson in the Higgs boson rest frame,
$\Omega_1^*$ and $\Omega_2^*$ are the decay solid angles of produced leptons in corresponding electroweak boson
rest frame, $\phi_1$ and $\phi_2$ are the azimuthal angles
of incoming off-mass shell gluons having the fractions $x_1$ and $x_2$ of the longitudinal momenta of 
colliding protons, $\lambda(x,y,z)$
is the kinematical function\cite{61}.
The cross section of the inclusive Higgs production in the diphoton decay mode 
can be written as\footnote{There was a missing factor $1/2$ in~(10) of\cite{52}, which
is due to identity of the final state photons. The numerical results\cite{52} have been corrected 
recently, conclusions unchanged.}:
\begin{equation}
  \displaystyle \sigma = {1\over 2\pi} \int {1 \over 16 \, x_1 x_2 s \, \lambda^{1/2}(\hat s, k_1^2, k_2^2) } \, f_g(x_1,{\mathbf k}_{1T}^2,\mu^2) f_g(x_2,{\mathbf k}_{2T}^2,\mu^2) |{\cal \bar M}^2| \, \times \atop {
  \displaystyle \times \, d{\mathbf k}_{1T}^2 d{\mathbf k}_{2T}^2 dy d\hat s d\Omega^* {d\phi_1 \over 2\pi} {d\phi_2 \over 2\pi}},
\end{equation}

\noindent
where $\Omega^*$ is the decay solid angle of produced photon in the Higgs boson rest frame.
This expession is more convenient for narrow Higgs resonance than the one used earlier\cite{52}.

Concerning the TMD gluon density functions in a proton, we have tested a few sets.
First of them (JH'2013 set 2) was obtained\cite{60} from the numerical solution of the 
CCFM equation. The latter seems to be the most suitable tool for our consideration
because it smoothly interpolates between the small-$x$ BFKL gluon dynamics 
and conventional DGLAP one, as it was mentioned above.
The input parameters of starting (initial) gluon distribution were fitted to 
describe the high-precision DIS data on proton structure functions $F_2(x,Q^2)$ and $F_2^c(x,Q^2)$\cite{60}.
The fit is based on TMD matrix elements and involves two-loop strong coupling constant,
kinematic consistency constraint\cite{62,63} and non-singular terms in the CCFM gluon splitting function\cite{64}.
Below we use this TMD gluon distribution as default 
choice\footnote{At the moment, there is a large variety of proposed TMD gluon distribution functions
in a proton. Most of them is collected in the \textsc{tmdlib} package\cite{65}, which is a C++ library 
providing a framework and an interface to the different parametrizations.}.
Additionaly, as an alternative choice, we apply the TMD gluon density obtained from the 
KMR prescription\cite{57}. The KMR approach is a formalism to construct
the TMD quark and gluon densities from well-known conventional ones. 
The key assumption of this approach is that the $k_T$-dependence of the TMD parton 
distributions enters at the last evolution step, so that the DGLAP evolution
can be used up to this step. 
For the input, we used Martin-Stirling-Thorn-Watt (MSTW'2008 LO) set\cite{66}.

Other essential parameters were taken as follows: the renormalization and factorization
scales $\mu_R^2 = \xi^2 m_H^2$ and $\mu_F^2 = \hat s + {\mathbf Q}_T^2$, where ${\mathbf Q}_T^2$
is the transverse momentum of the incoming off-shell gluon pair\footnote{The special choice for $\mu_F$ scale 
is connected with the CCFM evolution\cite{60}.}. 
To estimate the scale uncertainties of numerical 
calculations, we vary the unphysical parameter $\xi$ between $1/2$ and $2$ about the default value $\xi = 1$.
Following\cite{67}, we set electroweak bosons masses $m_Z = 91.1876$~GeV and $m_W = 80.403$~GeV, 
their total decay widths $\Gamma_Z = 2.4952$~GeV and $\Gamma_W = 2.085$~GeV. Additionally, 
we use Higgs boson mass $m_H = 126.8$~GeV, its full decay width $\Gamma_H = 4.3$~MeV, 
$\sin^2 \theta_W = 0.23122$ and adopt the LO formula for the strong coupling constant
$\alpha_s(\mu^2)$ with $n_f= 4$ active quark flavors at
$\Lambda_{\rm QCD} = 200$~MeV, so that $\alpha_s (m_Z^2) = 0.1232$.
Note that we use the running QED coupling
constant $\alpha(\mu^2)$.
Finally, following\cite{49}, to take into account the non-logarithmic loop 
corrections to the Higgs production cross section we apply the effective $K$-factor 
when using the KMR gluon density:
\begin{equation}
  K = \exp \left[ C_A {\alpha_s(\mu^2)\over 2\pi} \pi^2 \right],
\end{equation}

\noindent
where the color factor $C_A = 3$. A particular scale choice
$\mu^2 = {\mathbf p}_T^{4/3} {\hat s}^{2/3}$
(with ${\mathbf p}_T$ being the transverse momentum of
produced Higgs boson) has been proposed\cite{49}
to eliminate sub-leading logarithmic terms. We choose this scale
to evaluate the strong coupling constant in~(11) only.
The multidimensional integration everywhere was performed by means of a 
Monte Carlo technique, using the routine \textsc{vegas}\cite{68}.

\section{Numerical results} \indent

Now we are in a position to present our numerical results and discussion.
Let us consider first the Higgs boson production in the diphoton decay mode.

\subsection{$H \to \gamma \gamma$ decay mode} \indent

All cross sections were measured in a restricted part
of the phase space (fiducial phase space) defined to match 
the experimental acceptance in terms of the photon kinematics and topological 
event selection.
We implemented experimental setup used by the CMS and ATLAS Collaborations
in our numerical program.
In the CMS analysis\cite{11} performed at $\sqrt s = 8$~TeV two isolated photons
originating from the Higgs boson decays are required to have
pseudorapidities $|\eta^\gamma| < 2.5$. Additionally, photons with largest and 
next-to-largest transverse momentum $p_T^\gamma$ (so-called leading and subleading photons) 
must satisfy the conditions
of $p_T^\gamma/m^{\gamma \gamma} > 1/3$ and $p_T^\gamma/m^{\gamma \gamma} > 1/4$ respectively,
where $m^{\gamma \gamma}$ is the diphoton pair mass.
In the ATLAS measurement\cite{14} performed at $\sqrt s = 8$~TeV both of these decay photons must have 
pseudorapidities $|\eta^\gamma| < 2.37$ with the 
leading (subleading) photon satisfying $p_T^\gamma/m^{\gamma \gamma} > 0.35$~$(0.25)$, 
while invariant mass $m^{\gamma \gamma}$ is required to be $105 < m^{\gamma \gamma} < 160$~GeV.
The same kinematical cuts were applied in the preliminary measurements
performed by the CMS\cite{17} and ATLAS\cite{19} Collaborations at $\sqrt s = 13$~TeV with only
exception that invariant mass $m^{\gamma \gamma}$ in the CMS analysis\cite{17} should lie in the 
range $100 < m^{\gamma \gamma} < 180$~GeV. 
The diphoton pair transverse momentum $p_T^{\gamma \gamma}$, absolute value of the rapidity 
$|y^{\gamma \gamma}|$, photon helicity angle $\cos \theta^*$ (in the Collins-Soper frame)
and difference in azimuthal angle $\Delta \phi^{\gamma \gamma}$ between
the produced photons were measured\cite{11,14,17,19}.
Both $p_T^{\gamma \gamma}$ and $y^{\gamma \gamma}$ probe the 
production mechanism and parton distribution functions in a proton,
while $\cos \theta^*$ and $\Delta \phi^{\gamma \gamma}$ 
are related to properties (namely, spin-CP nature) of the decaying Higgs boson.

\begin{table}
\begin{center}
\begin{tabular}{|c|c|c|}
\hline
  & & \\
   Source  & $\sigma_{\rm fid}$(CMS) [fb] & $\sigma_{\rm fid}$(ATLAS) [fb] \\
  & & \\
\hline
  & & \\
  $k_T$-fact.,  JH'2013 set 2 & $31.12^{+4.71}_{-0.43}$ & $29.62^{+4.31}_{-0.32}$ \\
  & & \\
  $k_T$-fact., KMR & $22.47^{+11.98}_{-8.47}$ & $21.38^{+11.24}_{-8.01}$ \\
  & & \\
  fixed-order pQCD & $31^{+4}_{-3}$ & $30.5 \pm 3.3$ \\
  & & \\
\hline
  & & \\
  Measurement & $32 \pm 10$(stat.)$\pm 3$(syst.) & $43.2 \pm 9.4$(stat.)$^{+3.2}_{-2.9}$(syst.)$\pm 1.2$(lumi.) \\
  & & \\
\hline
\end{tabular}
\caption{The fiducial cross sections of inclusive Higgs boson production
(in the diphoton decay mode) at $\sqrt s = 8$~TeV. 
The experimental data are from CMS\cite{11} and ATLAS\cite{14}.
The results obtained in the collinear pQCD factorization (taken from\cite{11,14}) 
are shown for comparison.}
\label{table1}
\end{center}
\end{table}

\begin{table}
\begin{center}
\begin{tabular}{|c|c|c|}
\hline
  & & \\
   Source  & $\sigma_{\rm fid}$(CMS) [fb] & $\sigma_{\rm fid}$(ATLAS) [fb] \\
  & & \\
\hline
  & & \\
  $k_T$-fact., JH'2013 set 2 & $69.96^{+7.11}_{-0.53}$ & $68.23^{+6.69}_{-0.59}$ \\
  & & \\
  $k_T$-fact., KMR & $50.78^{+24.48}_{-17.99}$ & $47.91^{+23.59}_{-17.39}$ \\
  & & \\
  fixed-order pQCD & $75 \pm 4$ & $62.8^{+3.4}_{-4.4}$ \\
  & & \\
\hline
  & & \\
  Measurement & $84 \pm 11$(stat.)$\pm 7$(syst.) & $43.2 \pm 14.9$(stat.)$\pm 4.9$(syst.) \\
  & & \\
\hline
\end{tabular}
\caption{The fiducial cross sections of inclusive Higgs boson production
(in the diphoton decay mode) at $\sqrt s = 13$~TeV. 
The preliminary experimental data are from CMS\cite{17} and ATLAS\cite{19}.
The results obtained in the collinear pQCD factorization (taken from\cite{17,19}) 
are shown for comparison.}
\label{table2}
\end{center}
\end{table}

The results of our calculations are shown in Figs.~1 --- 3 
in comparison with the LHC data.
The solid histograms were obtained with the JH'2013 set 2 gluon density
by fixing both the renormalization $\mu_R$ and factorization $\mu_F$ scales at the
default values, while shaded regions
correspond to scale uncertainties of our predictions.
Following to\cite{60}, to estimate the latter
we used the JH'2013 set 2$+$ and JH'2013 set 2$-$ sets instead
of default one. These two sets 
represent a variation of the renormalization scale used in the 
off-shell production amplitude. The JH'2013 set 2$+$ set stands for a variation of $2\mu_R$, while
set JH'2013 set 2$-$ reflects $\mu_R/2$ (see also\cite{60} for more information).
One can see that the $k_T$-factorization predictions reasonably agree with the LHC
data within the experimental and theoretical uncertainties
for all considered kinematical
observables, although some tendency to slightly underestimate the ATLAS data (see Fig.~2)
and CMS data at large transverse momenta $p_T^{\gamma \gamma}$ (see Fig.~1) is observed
for both c.m. energies $\sqrt s = 8$ and $13$~TeV.
It could be due to the missing
contributions from the weak boson fusion ($W^+W^- \to H$ and $ZZ \to H$)
and/or associated $HZ$ or $HW^{\pm}$ production\cite{54}, 
which become important at high $p_T^{\gamma \gamma}$
and not taken into account in the present consideration.
Our results for $y^{\gamma \gamma}$ and $\cos\theta^*$ distributions obtained with the JH'2013 set 2 gluon
at $\sqrt s = 8$~TeV are consistently close to 
the matched NNLO + NNLL pQCD predictions obtained 
using the \textsc{hres} routine\cite{69} within the 
collinear QCD factorization (but a bit higher).
Our predictions at $\sqrt s = 13$~TeV are similar to the \textsc{nnlops} and a\textsc{mc@nlo}
ones\footnote{We take these predictions from the CMS\cite{11,14} and ATLAS\cite{17,19} papers.}.
It can be explained by the fact that the main part of collinear QCD higher-order corrections
(namely, NLO + NNLO + N$^3$LO + ... contributions which correspond to the $\log 1/x$ enhanced 
terms in perturbative series) are effectively taken into account as a part of the CCFM gluon 
evolution\footnote{The conventional high-order QCD corrections are known to be large, of about $80 - 100$\%\cite{6,7,8,9}.}.
A similar observation was done earlier\cite{47,49,52} and confirmed recently\cite{53}.
The calculations based o the alternative KMR gluon density also
tend to underestimate the ATLAS data at small $p_T^{\gamma \gamma}$, although they
describe well the CMS data and ATLAS data at high transverse momenta.
Moreover, we find that these predictions (mainly for distributions in $y^{\gamma \gamma}$ or $\cos \theta^*$) 
are generally similar to the lower uncertainty bounds of 
matched NNLO + NNLL (and \textsc{nnlops} or a\textsc{mc@nlo}) pQCD calculations.
This can be explained from the fact that
the KMR procedure absorbs only single gluon emission
at the last step of evolution (or, in other words, initial state
gluon emission closest to the produced Higgs boson), that corresponds 
to taking into account of $\ln 1/x$ enhanced NLO contributions only.
One can see that the shapes of $y^{\gamma \gamma}$ or $\cos \theta^*$ distributions
calculated using the CCFM-evolved and KMR gluon densities practically coincide and
therefore the difference between the JH'2013 set 2 and KMR 
predictions for these observables can
illustrate the role of conventional high-order contributions above the NLO level.
Here we demonstrate again the main advantage of the $k_T$-factorization approach,
which gives us the possibility to estimate the size of higher-order corrections
and reproduce in a straighforward manner the main features
of cumbersome fixed-order pQCD calculations.
In contrast, one can see that the shapes of $p_T^{\gamma \gamma}$
distributions predicted by the JH'2013 set 2 and KMR gluon densities
are very different from each other. Of course, it is not surprising since
the Higgs boson transverse momentum is strongly 
related to the initial gluon transverse momenta\cite{45,47,48,49,50,51,52,53}. 
The importance of this observable to distinguish between the different 
non-collinear evolution scenarios was pointed out\cite{47,52,53}.
Moreover, the difference in azimuthal angle $\Delta \phi^{\gamma \gamma}$
is also very sensitive to the initial gluon transverse momenta (see Fig.~1).
Such sensitivity is well-known and was demonstrated earlier for number of 
processes (see, for example,\cite{46} and references therein).
Thus, we confirm the previous conclusions\cite{47,52,53} that 
these observables can impose constraints on the TMD gluon densities of the proton.

The estimated Higgs boson fiducial cross sections at $\sqrt s = 8$ and $13$~TeV 
are listed in Tables~1 and~2 in comparison with the available data and
conventional high-order pQCD calculations performed using the \textsc{hres}\cite{69}, \textsc{nnlops}\cite{34,35} and 
a\textsc{mc@nlo}\cite{70} tools.
One can see that the $k_T$-factorization predictions are close to 
corresponding fixed-order collinear pQCD results
and agree well with the LHC data within the theoretical
and experimental uncertainties.
The scale dependence of the $k_T$-factorization predictions (especially obtained with the 
KMR gluon density)
is significant and exceeds the uncertainties of
conventional fixed-order pQCD calculations (which are about of $10 - 11$\%)\footnote{Note that 
scale uncertainties of the CCFM-based predictions are comparable with the ones of 
higher-order collinear pQCD calculations.}. However, it could be easily 
understood because only the tree-level LO hard scaterring 
amplitudes are involved.
Moreover, it was argued\cite{58} that amending the leading-logarithmic evolution with 
different kinematical constraints should lead to reasonable QCD 
predictions, although still formally only in leading 
logarithmic accuracy (see also\cite{46}).

\subsection{$H \to ZZ^* \to 4l$ and $H \to W^+W^- \to e^\pm \mu^\mp \nu \bar \nu$ decay channels} \indent

Now we turn to the $H \to ZZ^* \to 4l$ and $H \to W^+W^- \to e^\pm \mu^\mp \nu \bar \nu$ decay channels.
The data for the first of them come from the CMS\cite{12} and ATLAS Collaborations\cite{15}.
In the ATLAS analysis\cite{15} done at $\sqrt s = 8$~TeV 
only events with a four-lepton invariant mass $118 < m_{4l} < 129$~GeV are kept and 
each lepton (electron or muon) must satisfy transverse momentum cut
$p_T > 6$~GeV and be in the pseudorapidity range $|\eta| < 2.47$. The highest-$p_T$ lepton
in the quadruplet must have $p_T > 20$~GeV and the second (third) lepton in $p_T$ order
must satisfy $p_T > 15$($10$)~GeV. These leptons are required to be separated from each other
by $\Delta R = \sqrt{(\Delta \eta)^2 + (\Delta \phi)^2} > 0.1$($0.2$) when having the same (different)
lepton flavors. The invariant mass $m_{12}$ of the lepton pair closest to the $Z$ boson mass (leading pair)
is required to be $50 < m_{12} < 106$~GeV.
The subleading pair is chosen as the remaining lepton pair with invariant mass $m_{34}$ closest to the
$Z$ boson mass and satisfying the requirement $12 < m_{34} < 115$~GeV.
The CMS measurement\cite{12} performed at the same energy $\sqrt s = 8$~TeV 
requires at least four leptons in the event with at least
one lepton having $p_T > 20$~GeV, another lepton having $p_T > 10$~GeV and the remaining ones
having $p_T > 7$ and $5$~GeV respectively. All leptons must have the 
pseudorapidity $|\eta| < 2.4$, the leading pair invariant mass $m_{12}$ must be
$40 < m_{12} < 120$~GeV and subleading one should be $12 < m_{34} < 120$~GeV.
Finally, the four-lepton invariant mass $m_{4l}$ must satisfy $105 < m_{4l} < 140$~GeV cut.
Such cuts allow one to identify the decay leptons as originating from
different $Z$ bosons (real and virtual) and the interference effects in case 
of the production of identical leptons thus can be neglected\footnote{Incorrect identification 
is possible but happens only approximately in $5$\% of events\cite{15}.}.
Similar to the diphoton decay, the measurements are performed
in several observables related to the Higgs boson production and decay, namely the Higgs
transverse momentum $p_T^H$ and rapidity $|y^H|$,
invariant mass of the subleading lepton pair $m_{34}$ and cosine 
of the leading lepton pair decay angle $|\cos \theta^*|$ in the four-lepton rest frame with respect to
the beam axis. 
While the distributions in the $p_T^H$ and $|y^H|$ observables are sensitive to the 
production mechanism and gluon densities in a proton,
the distributions in the decay variables $m_{34}$ and $|\cos \theta^*|$
are sensitive to the Lagrangian structure of Higgs interaction (spin/CP quantum numbers and 
higher-dimensional operators).
In the ATLAS analysis\cite{16} performed at $\sqrt s = 8$~TeV for the $H \to W^+W^- \to e^\pm \mu^\mp \nu \bar \nu$ decay channel, 
events are selected from those with exactly one electron and one muon with opposite
charge, a dilepton invariant mass $10 < m_{ll} < 55$~GeV, 
azimuthal angle difference $\Delta \phi^{ll} < 1.8$ and missing transverse momentum
(which is produced by the two neutrinos from the $W$ boson decays) $p_T^{\rm miss} > 20$~GeV.
The leading lepton is required to have $p_T > 22$~GeV, the other one 
is required to have $p_T > 15$~GeV and both of them should be in the range $|\eta| < 2.47$.
The CMS analysis\cite{13} requires $p_T > 20$($10$)~GeV for the leading (subleading) leptons
with $|\eta| < 2.5$, lepton pair invariant mass $m_{ll} > 12$~GeV, their
transverse momentum $p_T^{ll} > 30$~GeV and invariant mass of the leptonic system
in the transverse plane $m_T^{ll\nu\nu} > 50$~GeV.
The differential cross sections were measured as 
functions of Higgs boson transverse momentum $p_T^H$ and 
absolute value of the dilepton rapidity $|y^{ll}|$.
The latter is highly correlated to the Higgs boson rapidity
$y^H$ which can not be reconstructed experimentally 
in the $H \to W^+W^- \to e^\pm \mu^\mp \nu \bar \nu$ final state.
Of course, all the experimental cuts listed above are 
taken into account in the numerical evaluations.
The preliminary data reported by the CMS\cite{18} and ATLAS\cite{20} Collaborations 
at $\sqrt s = 13$~TeV were obtained using similar analysis strategy.

\begin{table}
\begin{center}
\begin{tabular}{|c|c|c|}
\hline
  & & \\
   Source  & $\sigma_{\rm fid}$(CMS) [fb] & $\sigma_{\rm fid}$(ATLAS) [fb] \\
  & & \\
\hline
  & & \\
  $k_T$-fact., JH'2013 set 2 & $1.61^{+0.22}_{-0.01}$ & $1.58^{+0.23}_{-0.01}$ \\
  & & \\
  $k_T$-fact., KMR & $1.22^{+0.59}_{-0.42}$ & $1.20^{+0.58}_{-0.43}$ \\
  & & \\
  fixed-order pQCD & $1.15^{+0.12}_{-0.13}$ & $1.30 \pm 0.13$ \\
  & & \\
\hline
  & & \\
  Measurement & $1.11^{+0.41}_{-0.35}$(stat.)$^{+0.14}_{-0.10}$(syst.)$^{+0.08}_{-0.02}$(mod.) & $2.11^{+0.53}_{-0.47}$(stat.)$\pm 0.08$(syst.) \\
  & & \\
\hline
\end{tabular}
\caption{The fiducial cross sections of inclusive Higgs production
(in the $H \to ZZ^* \to 4l$ decay channel) at $\sqrt s = 8$~TeV. 
The experimental data are from CMS\cite{12} and ATLAS\cite{15}.
The results obtained in the collinear pQCD factorization (taken from\cite{12,15}) 
are shown for comparison.}
\label{table3}
\end{center}
\end{table}

\begin{table}
\begin{center}
\begin{tabular}{|c|c|c|}
\hline
  & & \\
   Source  & $\sigma_{\rm fid}$(CMS) [fb] & $\sigma_{\rm fid}$(ATLAS) [fb] \\
  & & \\
\hline
  & & \\
  $k_T$-fact., JH'2013 set 2 & $54.47^{+8.20}_{-0.46}$ & $34.02^{+5.58}_{-0.38}$ \\
  & & \\
  $k_T$-fact., KMR & $40.80^{+21.33}_{-15.06}$ & $27.38^{+13.07}_{-9.39}$ \\
  & & \\
  fixed-order pQCD & $48 \pm 8$ & $25.1 \pm 2.6$ \\
  & & \\
\hline
  & & \\
  Measurement & $39 \pm 8$(stat.)$\pm 9$(syst.) & $36.0 \pm 7.2$(stat.)$\pm 6.4$(syst.)$\pm 1.0$(lumi.) \\
  & & \\
\hline
\end{tabular}
\caption{The fiducial cross sections of inclusive Higgs production
(in the $H \to W^+W^- \to e^\pm \mu^\mp \nu \bar \nu$ decay channel) at $\sqrt s = 8$~TeV. 
The experimental data are from CMS\cite{13} and ATLAS\cite{16}.
The results obtained in the collinear pQCD factorization (taken from\cite{13,16}) 
are shown for comparison.}
\label{table4}
\end{center}
\end{table}

\begin{table}
\begin{center}
\begin{tabular}{|c|c|c|}
\hline
  & & \\
   Source  & $\sigma_{\rm fid}$(CMS) [fb] & $\sigma_{\rm fid}$(ATLAS) [fb] \\
  & & \\
\hline
  & & \\
  $k_T$-fact., JH'2013 set 2 & $3.61^{+0.33}_{-0.01}$ & $3.84^{+0.38}_{-0.02}$ \\
  & & \\
  $k_T$-fact., KMR & $2.71^{+1.17}_{-0.90}$ & $2.83^{+1.28}_{-0.96}$ \\
  & & \\
  fixed-order pQCD & $2.76 \pm 0.14$ & $2.91 \pm 0.13$ \\
  & & \\
\hline
  & & \\
  Measurement & $2.92^{+0.48}_{-0.44}$(stat.)$^{+0.28}_{-0.24}$(syst.) & $3.62^{+0.53}_{-0.50}$(stat.)$^{+0.25}_{-0.20}$(syst.) \\
  & & \\
\hline
\end{tabular}
\caption{The fiducial cross sections of inclusive Higgs production
(in the $H \to ZZ^* \to 4l$ decay channel) at $\sqrt s = 13$~TeV. 
The preliminary experimental data are from CMS\cite{18} and ATLAS\cite{20}.
The results obtained in the collinear pQCD factorization (taken from\cite{18,20}) 
are shown for comparison.}
\label{table5}
\end{center}
\end{table}

The results of our calculations are shown in Figs.~4 --- 8 
in comparison with the data. The estimated
total cross sections are listed in Tables~3 --- 5.
Similar to $H \to \gamma \gamma$ decay, the $k_T$-factorization predictions 
for $H \to ZZ^* \to 4l$ and $H \to W^+W^- \to e^\pm \mu^\mp \nu \bar \nu$ decay modes
agree well with the LHC data taken at $\sqrt s = 8$~TeV for all considered kinematical
observables within the theoretical and experimental uncertainties.
The best description of the data is achieved with the CCFM-evolved JH'2013 set 2 gluon density.
Moreover, the overall agreement between these predictions
and the preliminary ATLAS data\cite{20} taken at $\sqrt s = 13$~TeV 
looks to be even 
a bit better then the one given by the NNLO pQCD calculations (see Fig.~8),
that could be essentially due to the small-$x$ region probed.
The KMR approach results in lower cross sections 
compared to the JH'2013 set 2 calculations since only single gluon emission in the initial state
is taken into account here.
Good agreement is also observed in the normalized differential cross sections
$1/\sigma \, d\sigma/dp_T^H$ and $1/\sigma \, d\sigma/d|y^{ll}|$ (see Fig.~7).
Studying of the normalized differential cross sections leads to a more 
stringent comparison between data and theory due to reduced experimental (mainly systematic)
uncertainties.
As it was expected, the distributions on the Higgs boson transverse momentum are highly sensitive to
the TMD gluon densities applied in the numerical calculations
and therefore can be used to discriminate between the latter.
In contrast, the predicted shapes of rapidity and $\cos\theta^*$ distributions 
are almost insensitive to the TMD gluon density in a proton.
The KMR predictions for these distributions
are rather similar to the lower uncertainty bounds of the
NNLOPS calculations, whereas the JH'2013 set 2 ones slightly overshoot them.
This fact demonstrates again the role of $\ln 1/x$-enhanced NNLO + N$^3$LO + ...
terms taken into account in the CCFM gluon evolution.

Finally, we would like to note that 
a similar study (but using the $H \to ZZ^* \to 4l$ decay channel only) was done very recently\cite{53}.
Unlike our choice, older version of CCFM-evolved gluon density in a proton (namely, set A0)\cite{72} was 
applied in these calculations. We reproduce the results\cite{53} when using the A0 gluon.

\section{Conclusions} \indent

We investigated the inclusive Higgs boson production in $pp$ collisions
at the LHC using the $H \to \gamma \gamma$, $H \to ZZ^* \to 4l$ and 
$H \to W^+W^- \to e^\pm \mu^\mp \nu \bar \nu$ decay channels 
in the framework of the $k_T$-factorization approach.
Our consideration was based 
on the dominant off-shell gluon-gluon fusion subprocess where 
the transverse momenta of initial gluons are taken into account.
The essential part of our analysis was using of the TMD gluon density 
derived from the CCFM evolution equation. 
The latter seems to be the most suitable tool for our consideration
because it smoothly interpolates between the small-$x$ BFKL gluon dynamics 
and conventional DGLAP one, which is valid at large Bjorken $x$.
Using the CCFM-evolved gluon density, we have achieved reasonably good description of the latest data 
taken by the CMS and ATLAS Collaborations at $\sqrt s = 8$~TeV and
recent preliminary data taken at $\sqrt s = 13$~TeV.
The theoretical uncertainties of our
calculations were estimated and comparison 
with the high-order pQCD predictions (up to NNLO + NNLL level) 
obtained within the collinear factorization was done. 
We have illustrated the effect of taking into account $\ln 1/x$-enhanced higher-order 
terms in our calculations and demonstrated the strong sensitivity 
of predicted Higgs transverse momentum distributions to the TMD gluon densities used.
Such observables could impose constraints on the latter.

\section{Acknowledgements} \indent

We would like to thank S.P.~Baranov, H.~Jung, V.~Rawoot and R.~Islam for their very useful 
discussions and important remarks. 
This research was supported in part by RFBR grant 16-32-00176-mol-a and
grant of the President of Russian Federation NS-7989.2016.2.
We are grateful to DESY Directorate for the 
support in the framework of Moscow --- DESY project on Monte-Carlo implementation for
HERA --- LHC. M.A.M. was also supported by a grant of the foundation for
the advancement of theoretical physics "Basis" 17-14-455-1.


\newpage 

\begin{figure}
\begin{center}
\includegraphics[width=8cm]{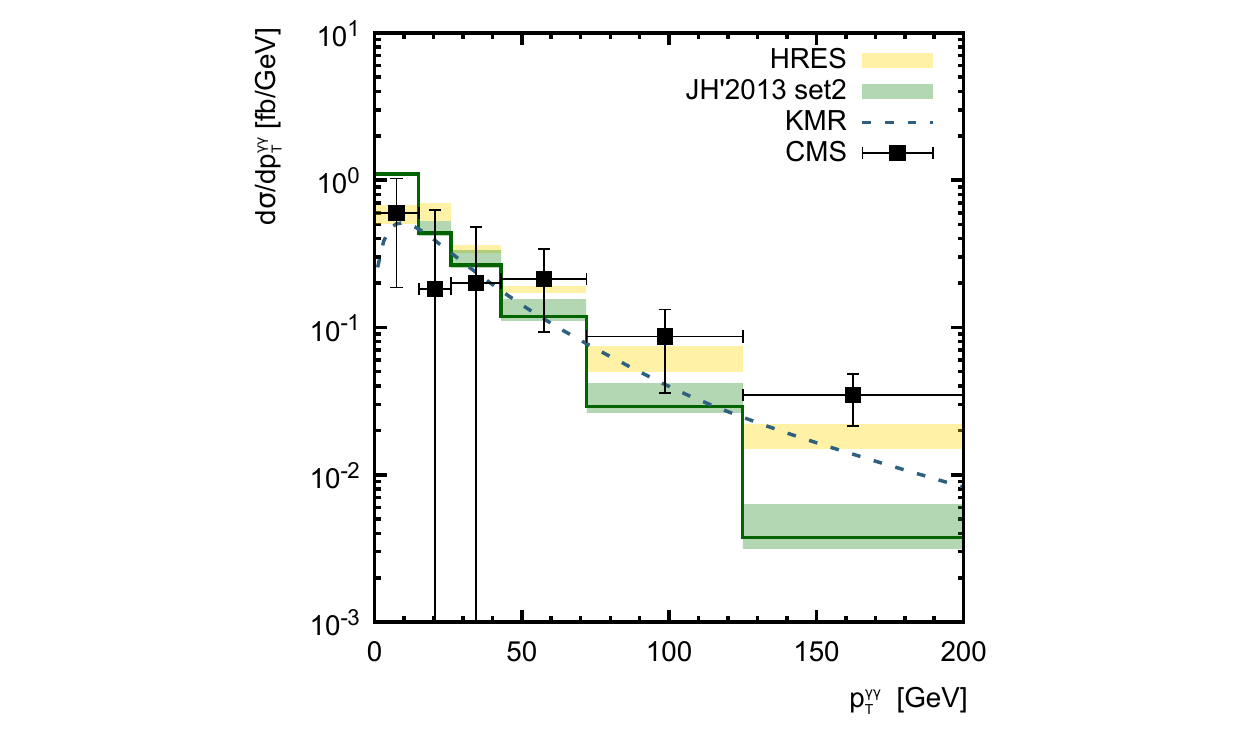}
\includegraphics[width=8cm]{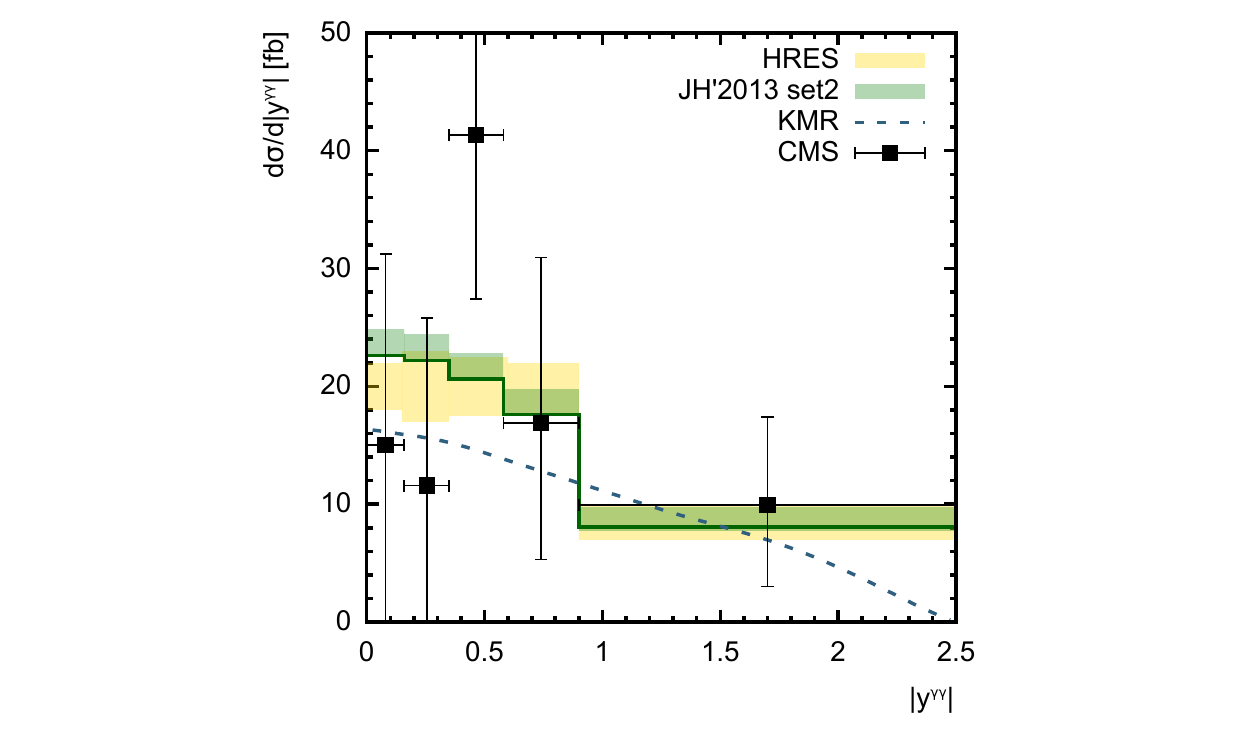}
\includegraphics[width=8cm]{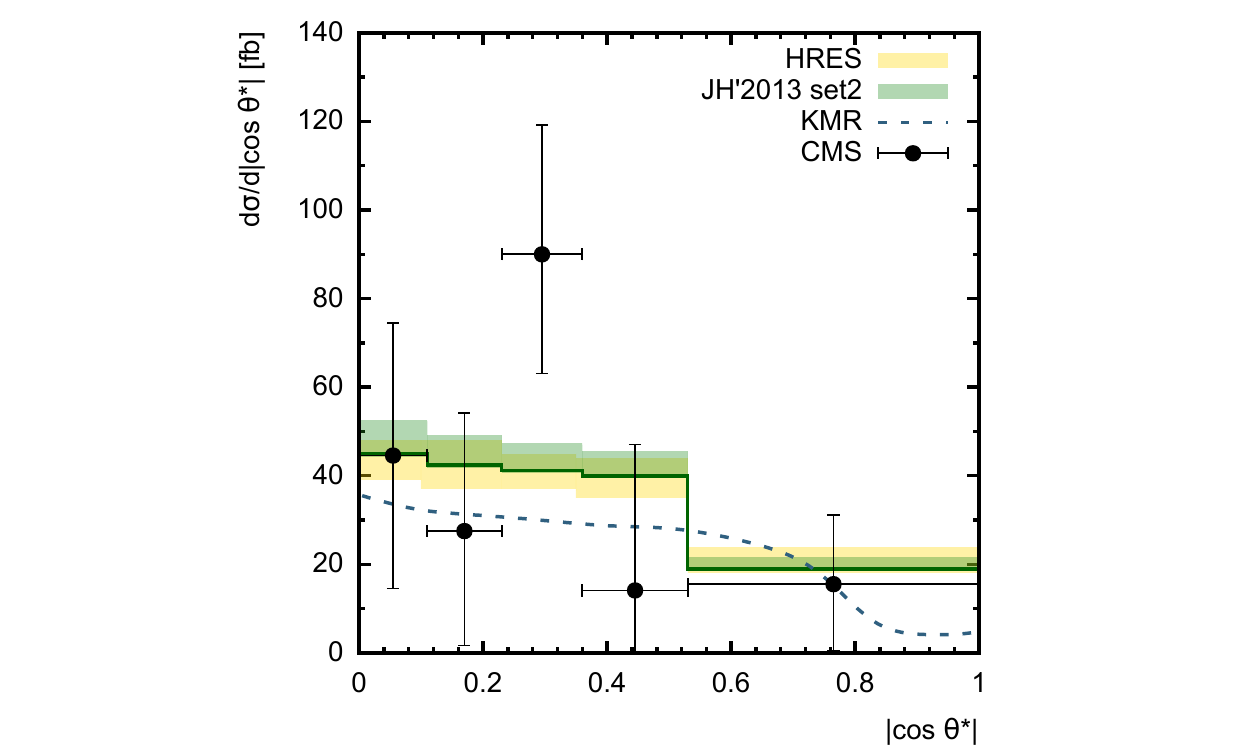}
\includegraphics[width=8cm]{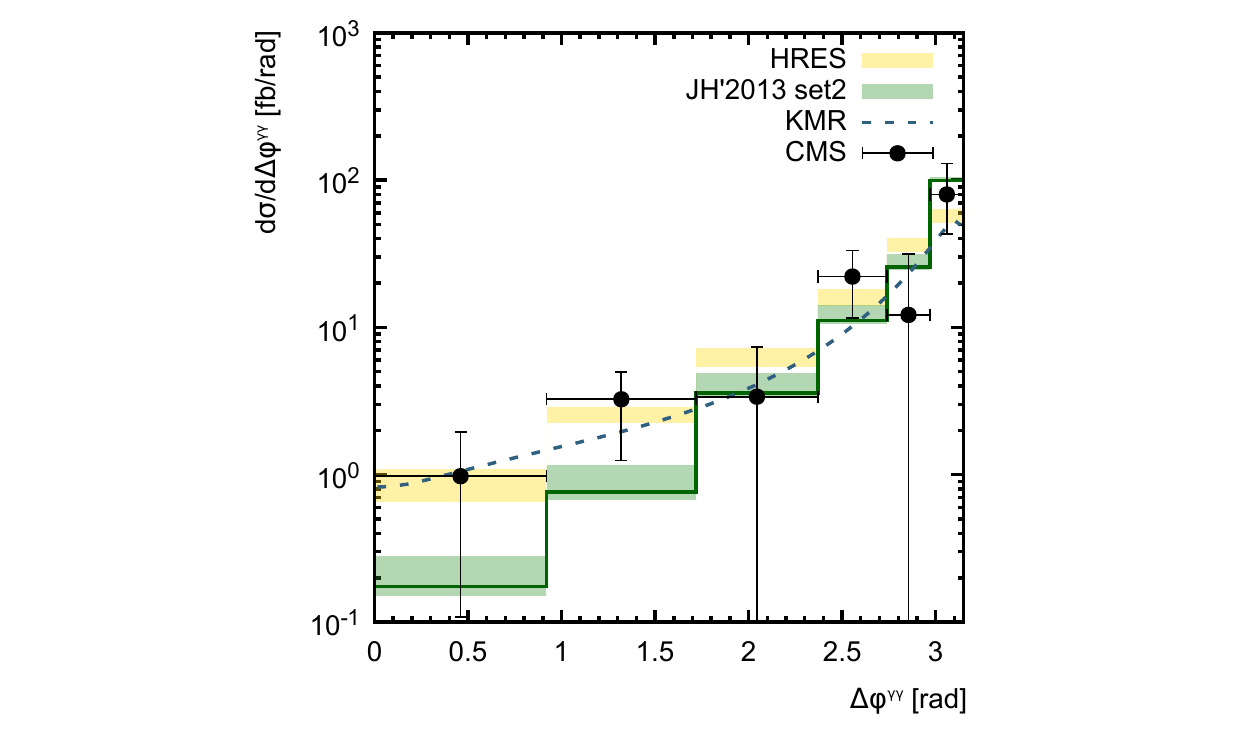}
\caption{The differential cross sections of inclusive Higgs boson 
production (in the diphoton decay mode) at $\sqrt s = 8$~TeV
as functions of diphoton pair transverse momentum $p_T^{\gamma \gamma}$, rapidity $|y^{\gamma \gamma}|$, 
azimuthal angle difference $\Delta \phi^{\gamma \gamma}$ and photon helicity angle $\cos \theta^*$ (in the Collins-Soper frame). 
The solid histograms represent the $k_T$-factorization predictions obtained with the JH'2013 set 2 gluon density
at the default hard scales. The shaded bands (green) represent the scale uncertainties of these calculations, as it is 
described in the text. The dashed curves correspond to the calculations with the KMR gluon density. 
The NNLO + NNLL pQCD predictions obtained using the \textsc{hres} routine\cite{69} (taken from\cite{11})
are presented as a hatched (blue) band. 
The experimental data are from CMS\cite{11}.}
\label{fig1}
\end{center}
\end{figure}

\begin{figure}
\begin{center}
\includegraphics[width=8cm]{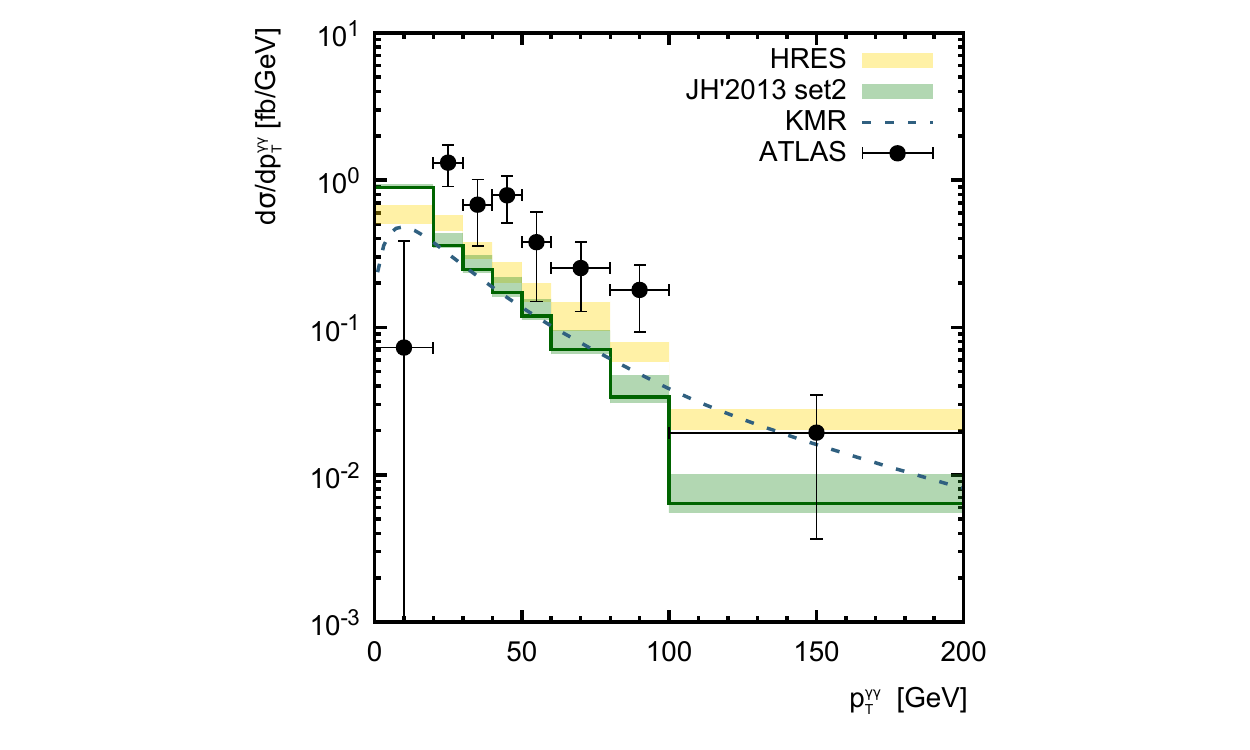}
\includegraphics[width=8cm]{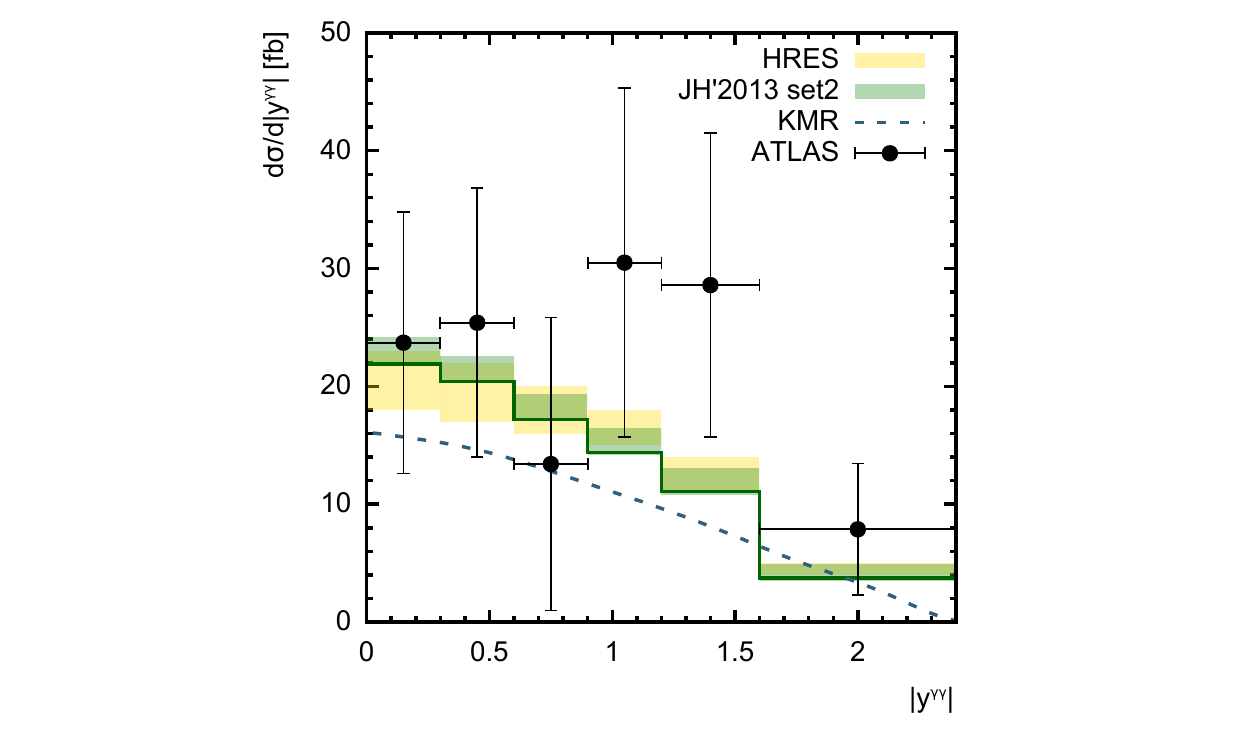}
\includegraphics[width=8cm]{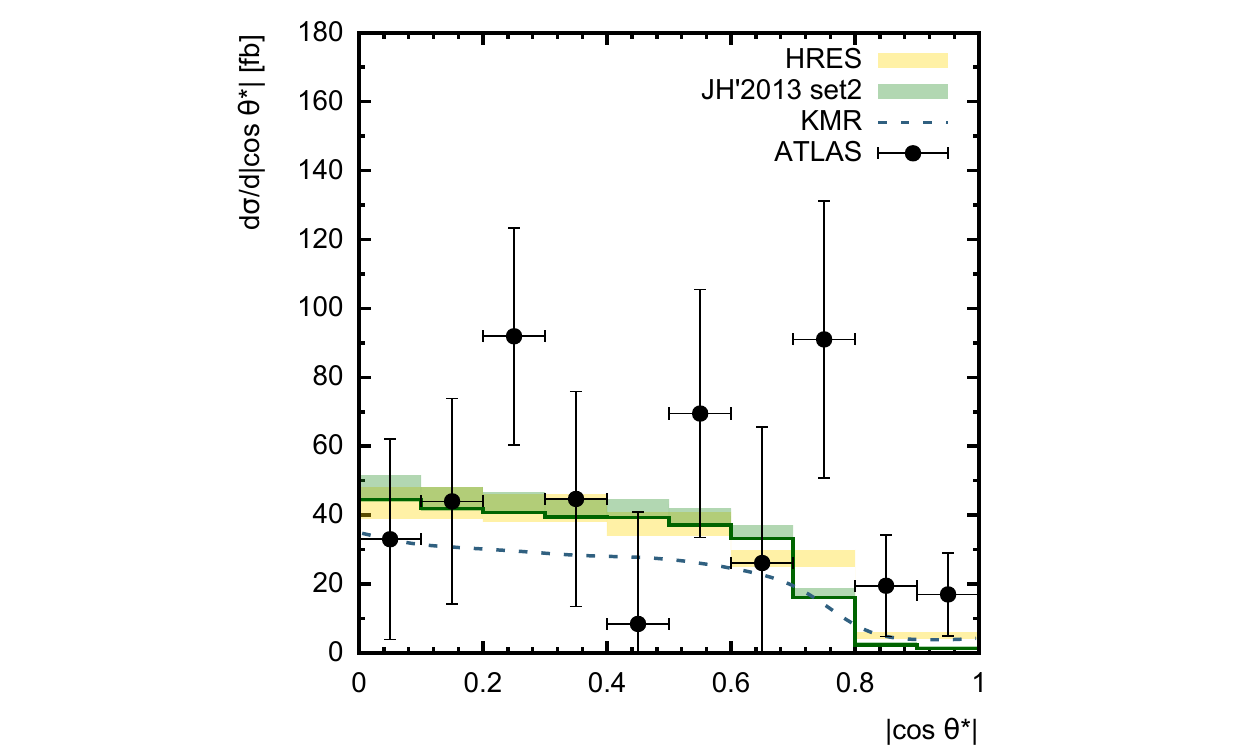}
\caption{The differential cross sections of inclusive Higgs boson 
production (in the diphoton decay mode) at $\sqrt s = 8$~TeV
as functions of diphoton pair transverse momentum $p_T^{\gamma \gamma}$, rapidity $|y^{\gamma \gamma}|$ 
and photon helicity angle $\cos \theta^*$ in the Collins-Soper frame. 
Notation of histograms and curves is the same as in Fig.~1. 
The experimental data are from ATLAS\cite{14}.
The \textsc{hres}\cite{69} predictions are taken from\cite{14}.}
\label{fig2}
\end{center}
\end{figure}

\begin{figure}
\begin{center}
\includegraphics[width=8cm]{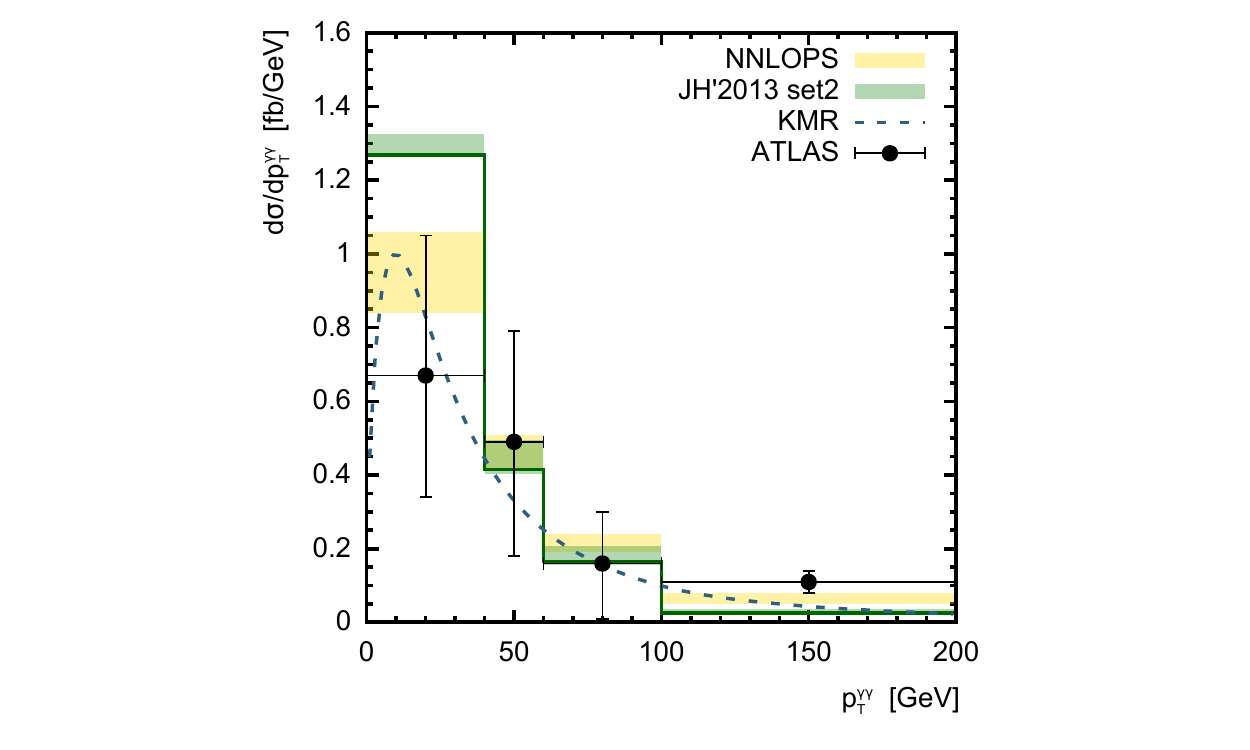}
\includegraphics[width=8cm]{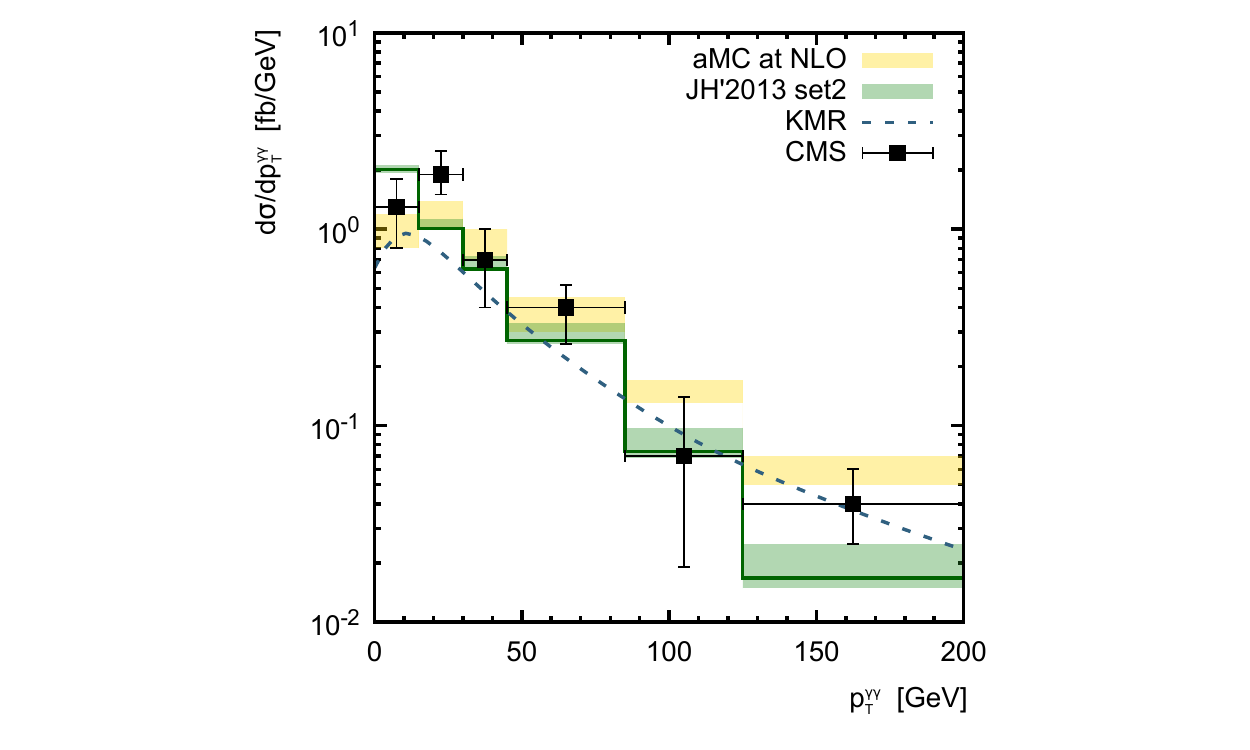}
\includegraphics[width=8cm]{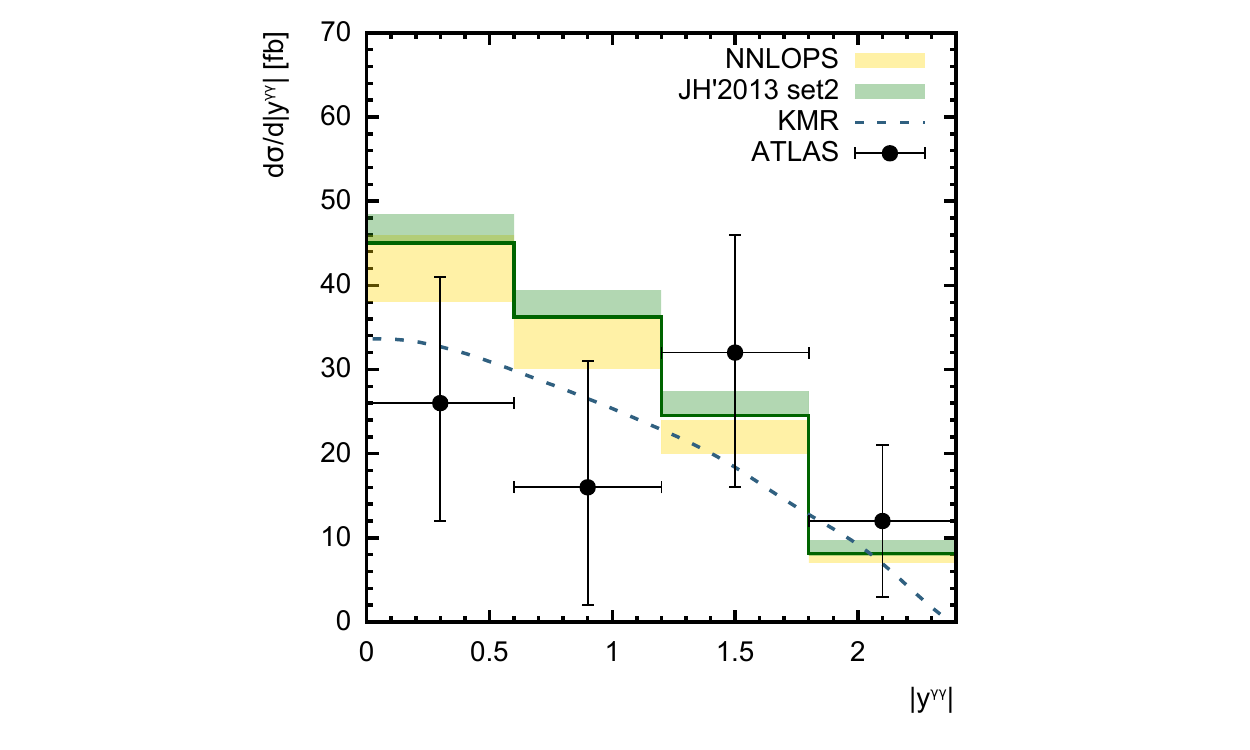}
\includegraphics[width=8cm]{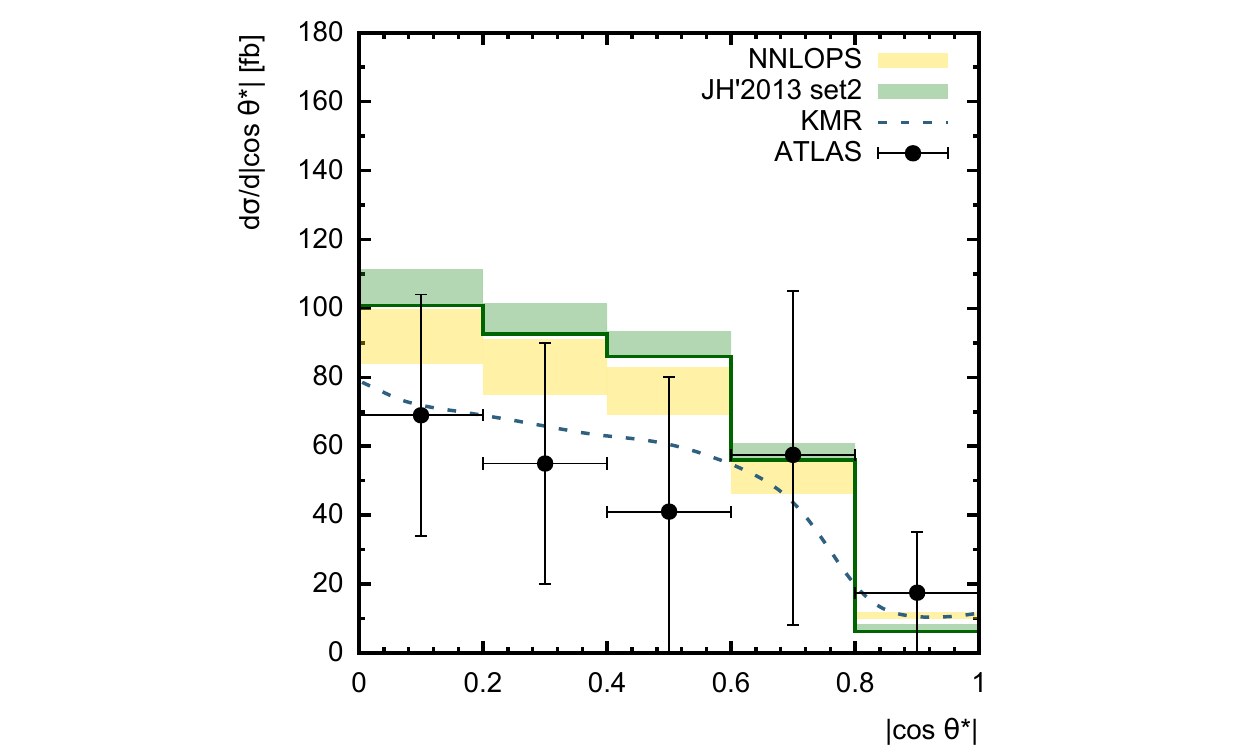}
\caption{The differential cross sections of inclusive Higgs boson 
production (in the diphoton decay mode) at $\sqrt s = 13$~TeV
as functions of diphoton pair transverse momentum $p_T^{\gamma \gamma}$, rapidity $|y^{\gamma \gamma}|$ 
and photon helicity angle $\cos \theta^*$ in the Collins-Soper frame. 
Notation of histograms and curves is the same as in Fig.~1. 
The preliminary experimental data are from CMS\cite{17} and ATLAS\cite{19}. 
The \textsc{nnlops}\cite{34,35} and a\textsc{mc@nlo}\cite{70} predictions are taken from\cite{17,19}.}
\label{fig3}
\end{center}
\end{figure}

\begin{figure}
\begin{center}
\includegraphics[width=8cm]{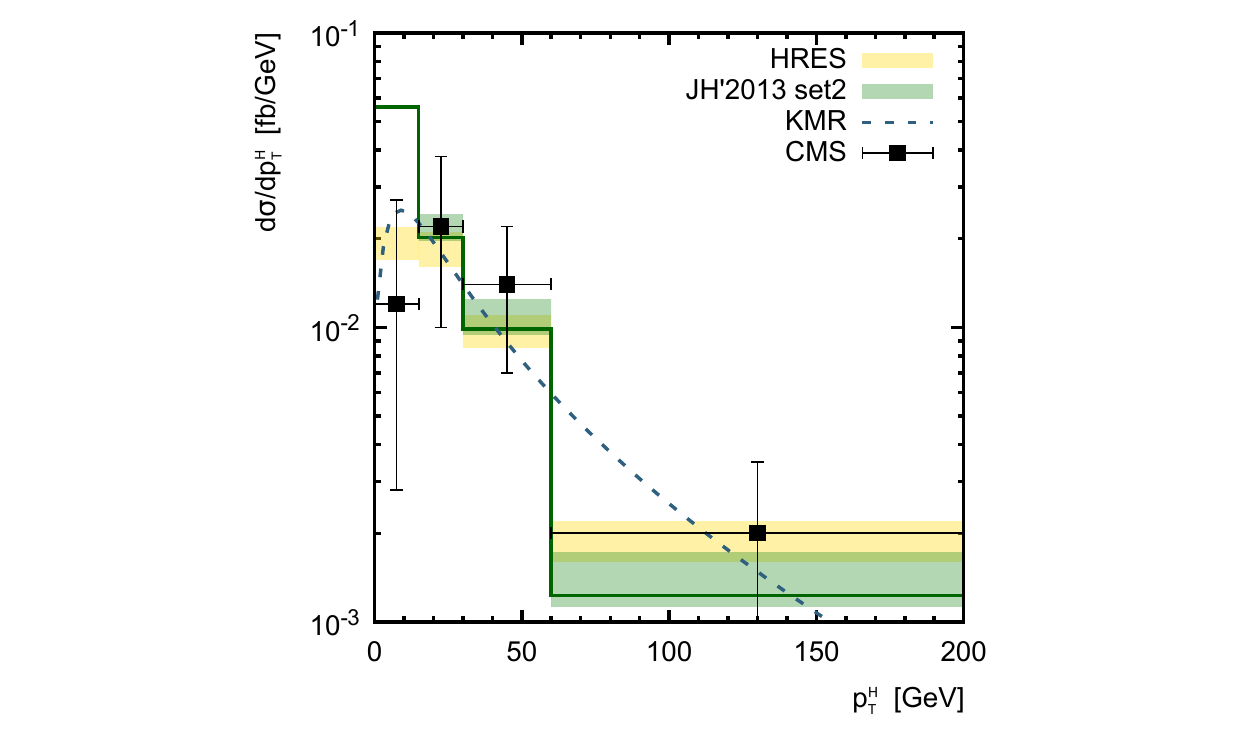}
\includegraphics[width=8cm]{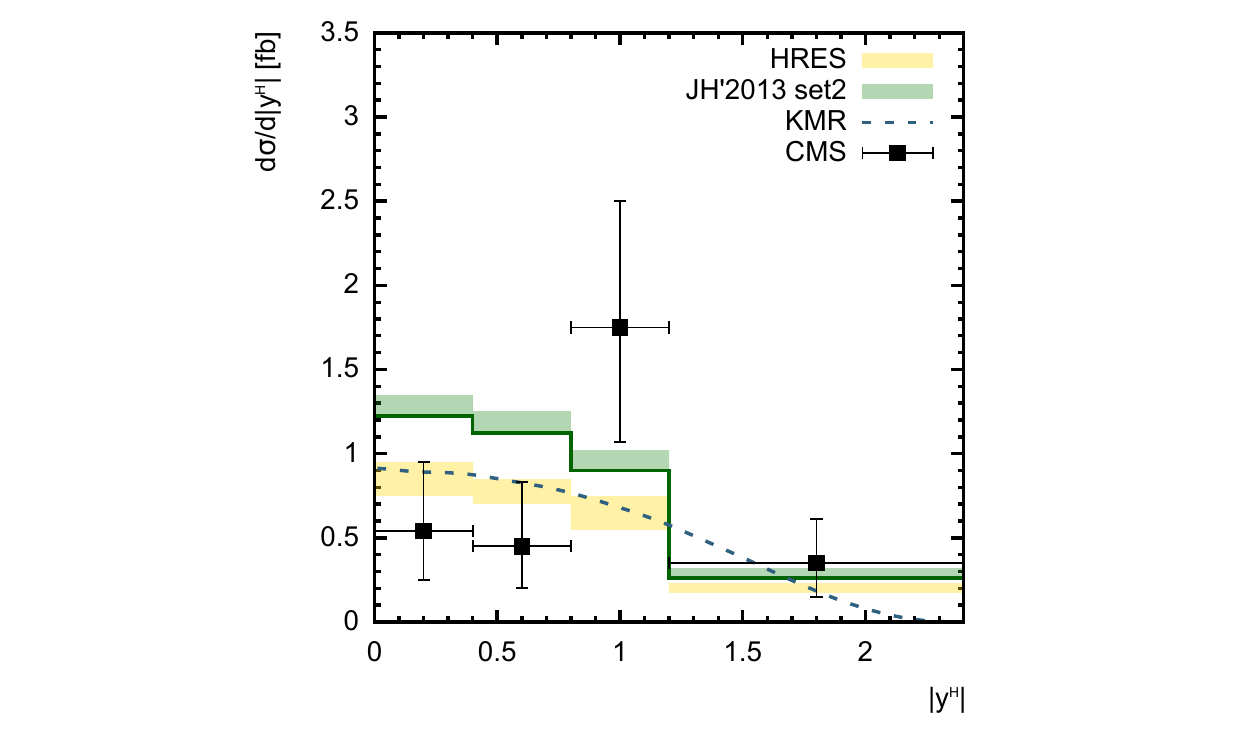}
\caption{The differential cross sections of inclusive Higgs boson 
production (in the $H \to ZZ^* \to 4l$ decay mode) at $\sqrt s = 8$~TeV
as functions of Higgs transverse momentum and rapidity. 
Notation of histograms and curves is the same as in Fig.~1. 
The experimental data are from CMS\cite{12}. 
The \textsc{hres}\cite{69} predictions are taken from\cite{12}.}
\label{fig4}
\end{center}
\end{figure}

\begin{figure}
\begin{center}
\includegraphics[width=8cm]{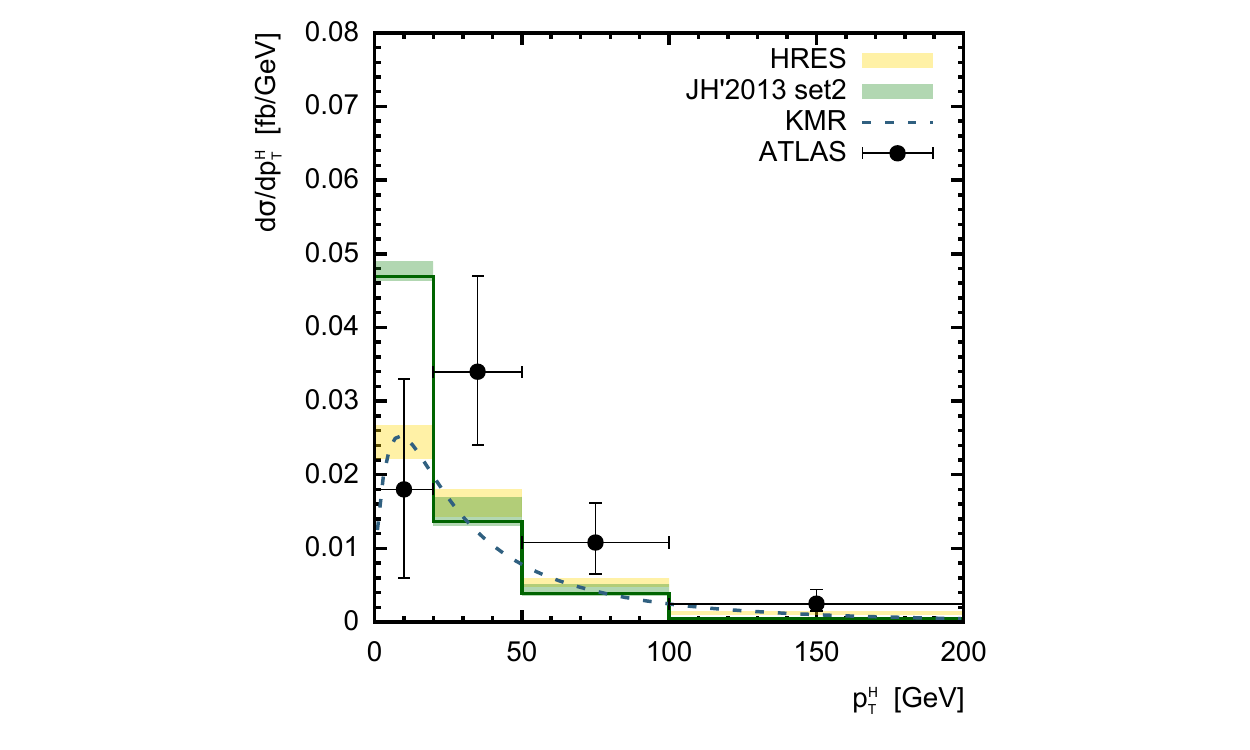}
\includegraphics[width=8cm]{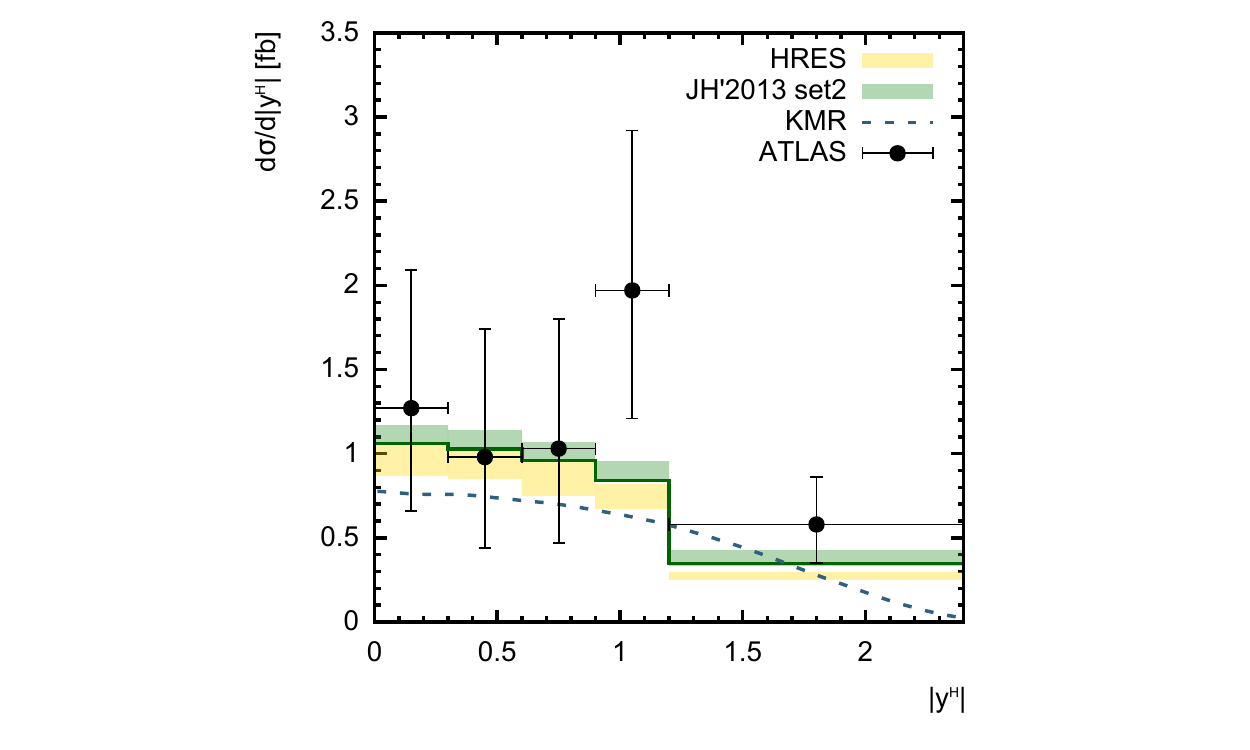}
\includegraphics[width=8cm]{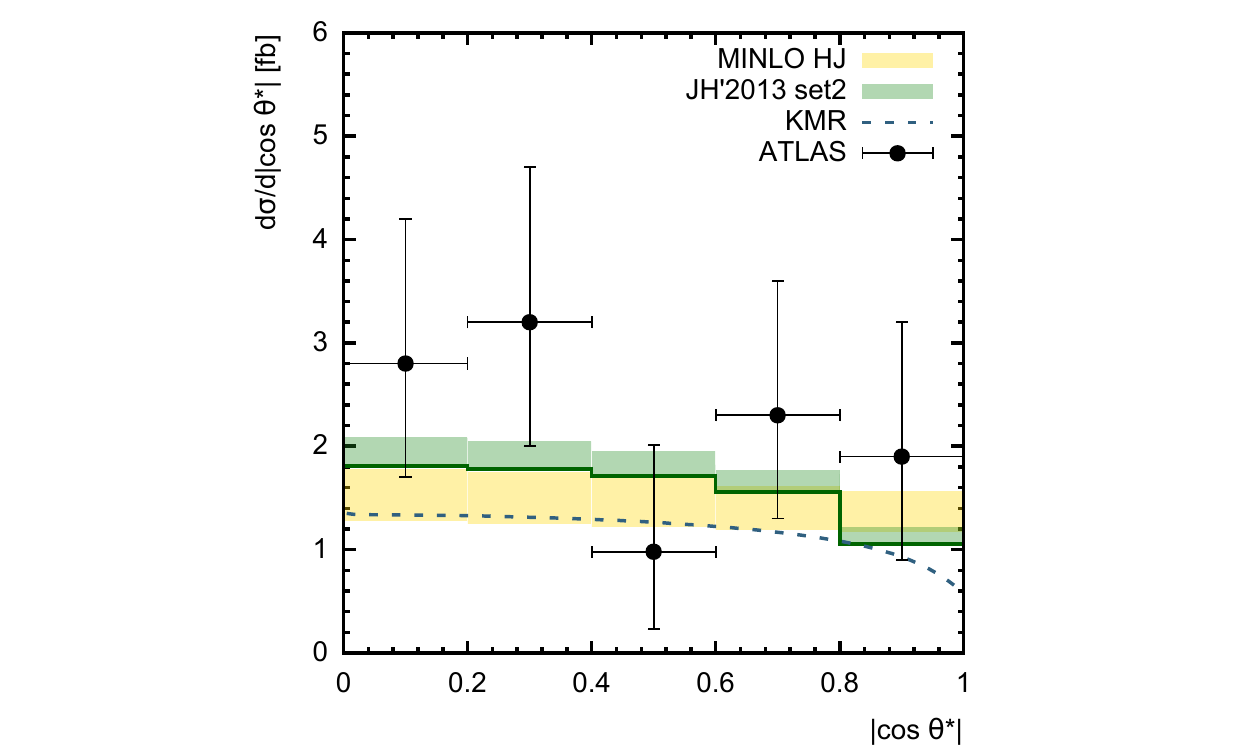}
\includegraphics[width=8cm]{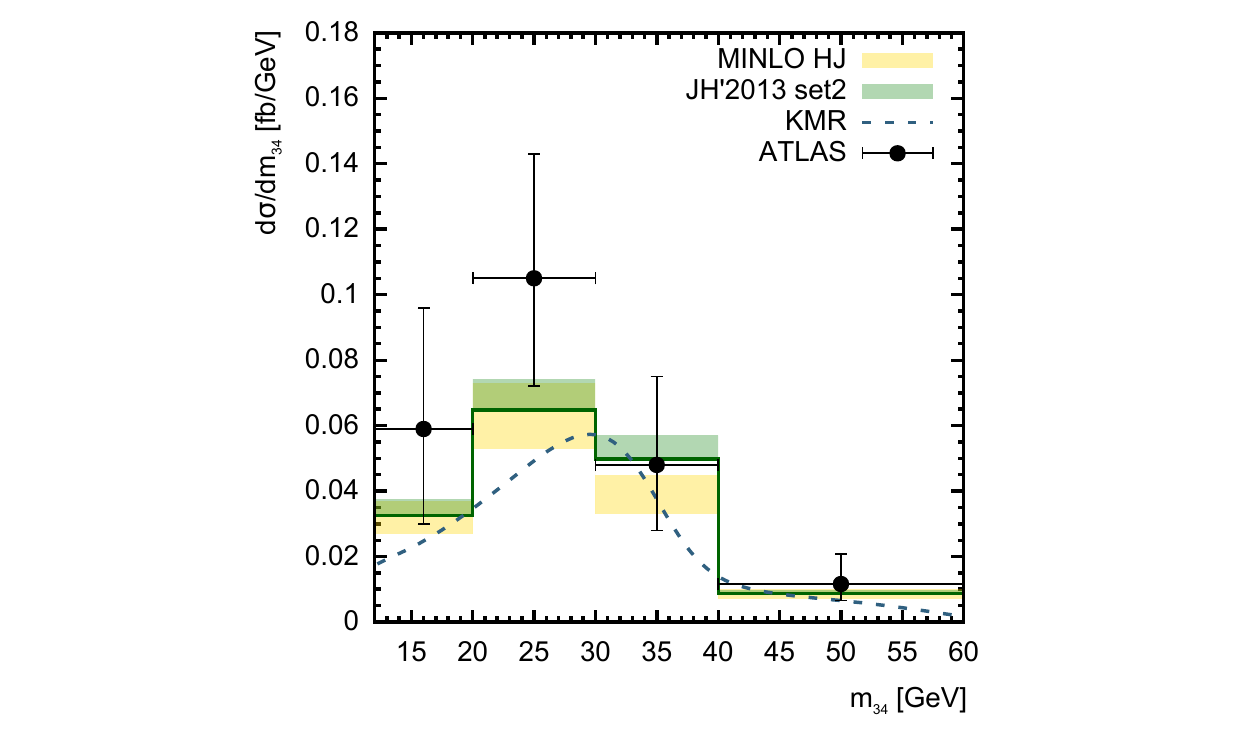}
\caption{The differential cross sections of inclusive Higgs boson 
production (in the $H \to ZZ^* \to 4l$ decay mode) at $\sqrt s = 8$~TeV
as functions of Higgs transverse momentum $p_T^{H}$, rapidity $|y^{H}|$, 
leading lepton pair decay angle $\cos \theta^*$ (in the Collins-Soper frame) and invariant mass $m_{34}$
of subleading lepton pair. 
Notation of histograms and curves is the same as in Fig.~1. 
The experimental data are from ATLAS\cite{15}.
The \textsc{hres}\cite{69} and \textsc{minlo hj}\cite{71} predictions are taken from\cite{15}.}
\label{fig5}
\end{center}
\end{figure}

\begin{figure}
\begin{center}
\includegraphics[width=8cm]{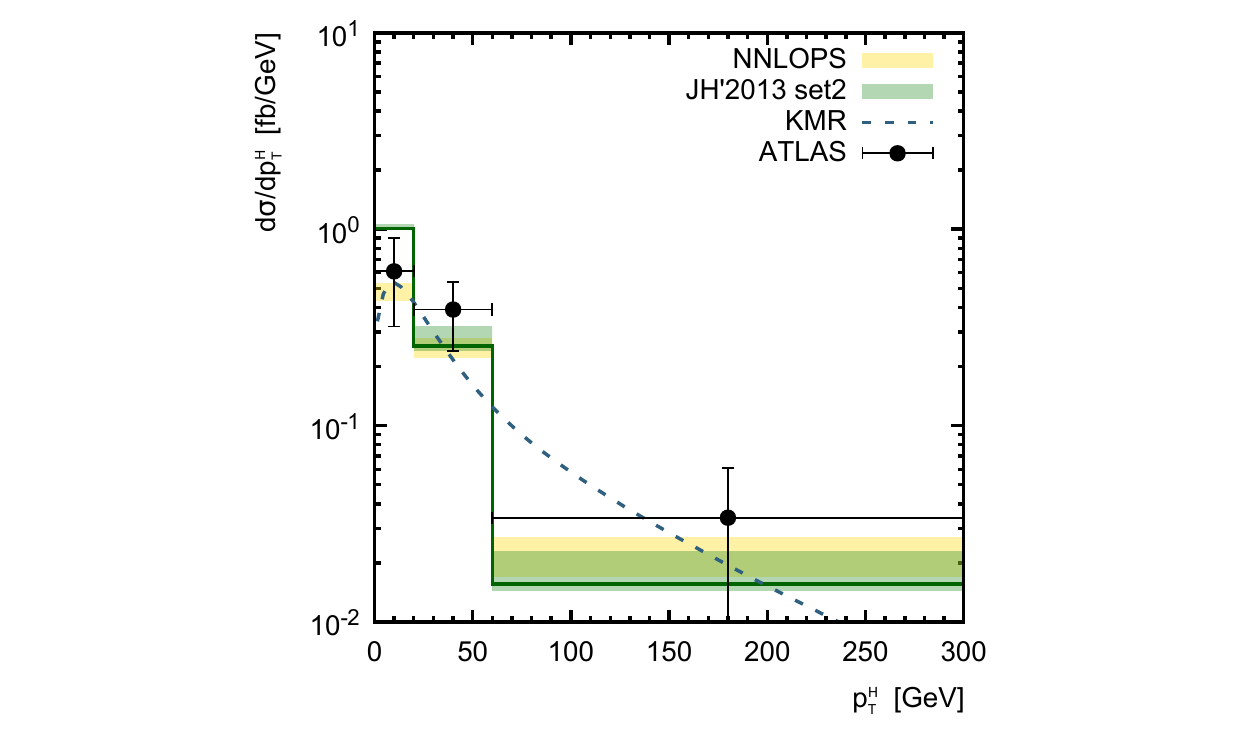}
\includegraphics[width=8cm]{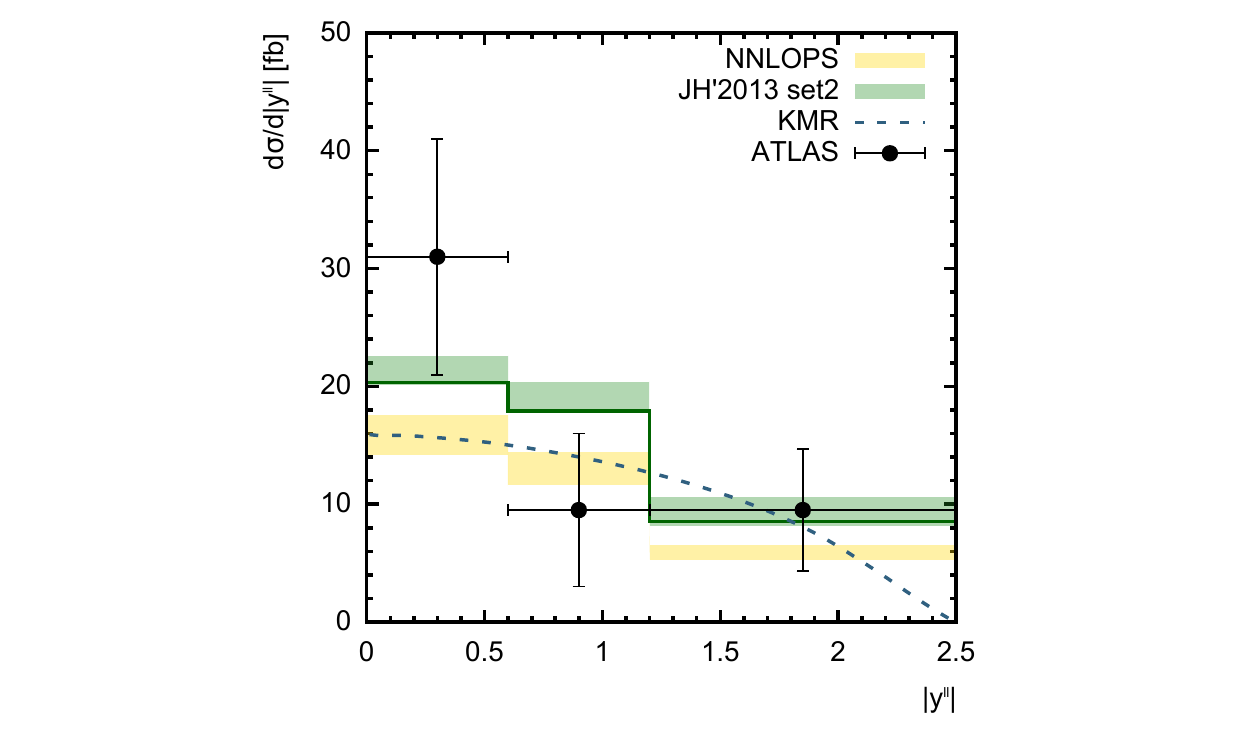}
\includegraphics[width=8cm]{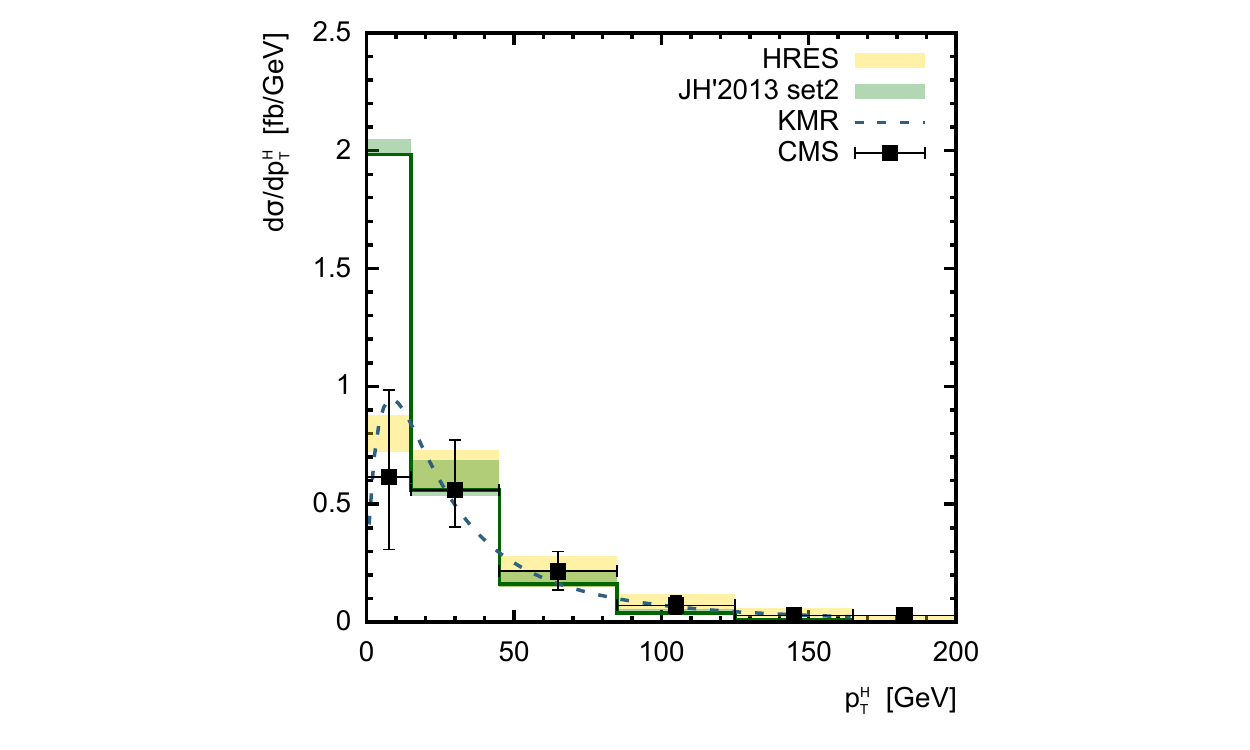}
\caption{The differential cross sections of inclusive Higgs  
production (in the $H \to W^+W^- \to e^\pm \mu^\mp \nu \bar \nu$ decay mode) at $\sqrt s = 8$~TeV
as functions of Higgs transverse momentum and lepton pair rapidity. 
Notation of histograms and curves is the same as in Fig.~1. 
The experimental data are from CMS\cite{13} and ATLAS\cite{16}. 
The \textsc{hres}\cite{69} and \textsc{nnlops}\cite{34,35} predictions are taken from\cite{13,16}.}
\label{fig6}
\end{center}
\end{figure}

\begin{figure}
\begin{center}
\includegraphics[width=8cm]{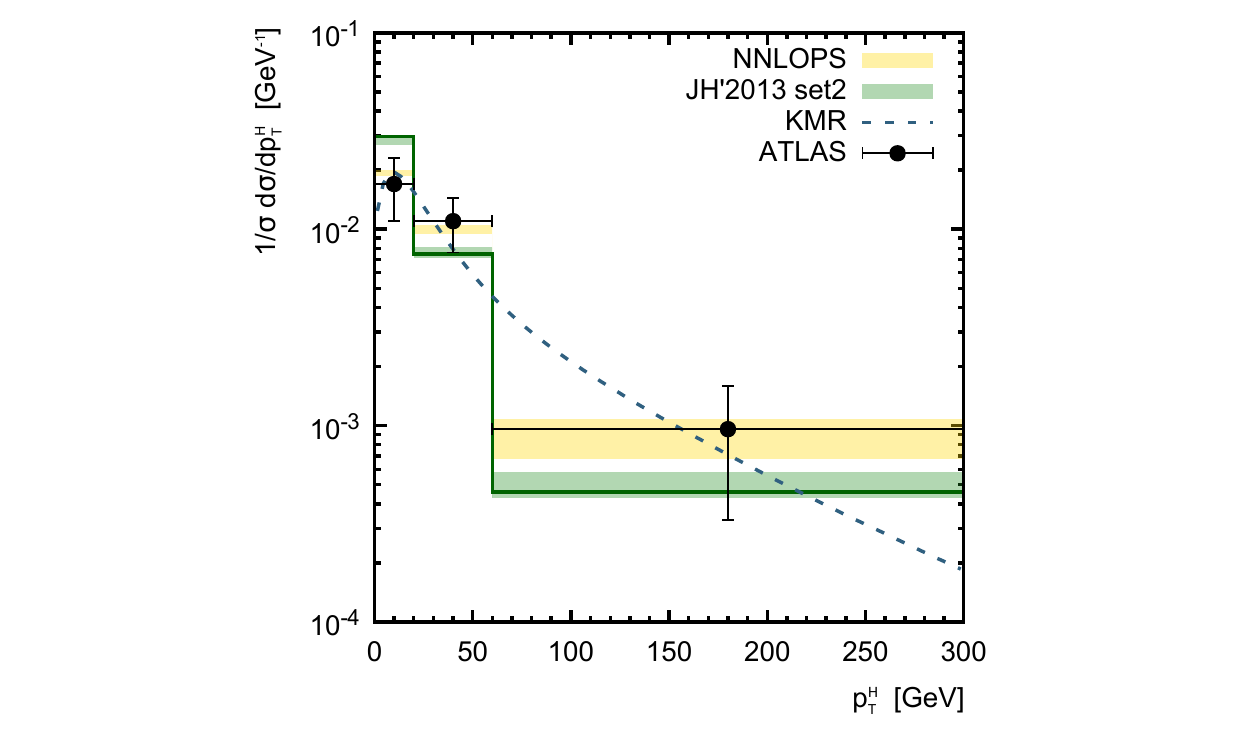}
\includegraphics[width=8cm]{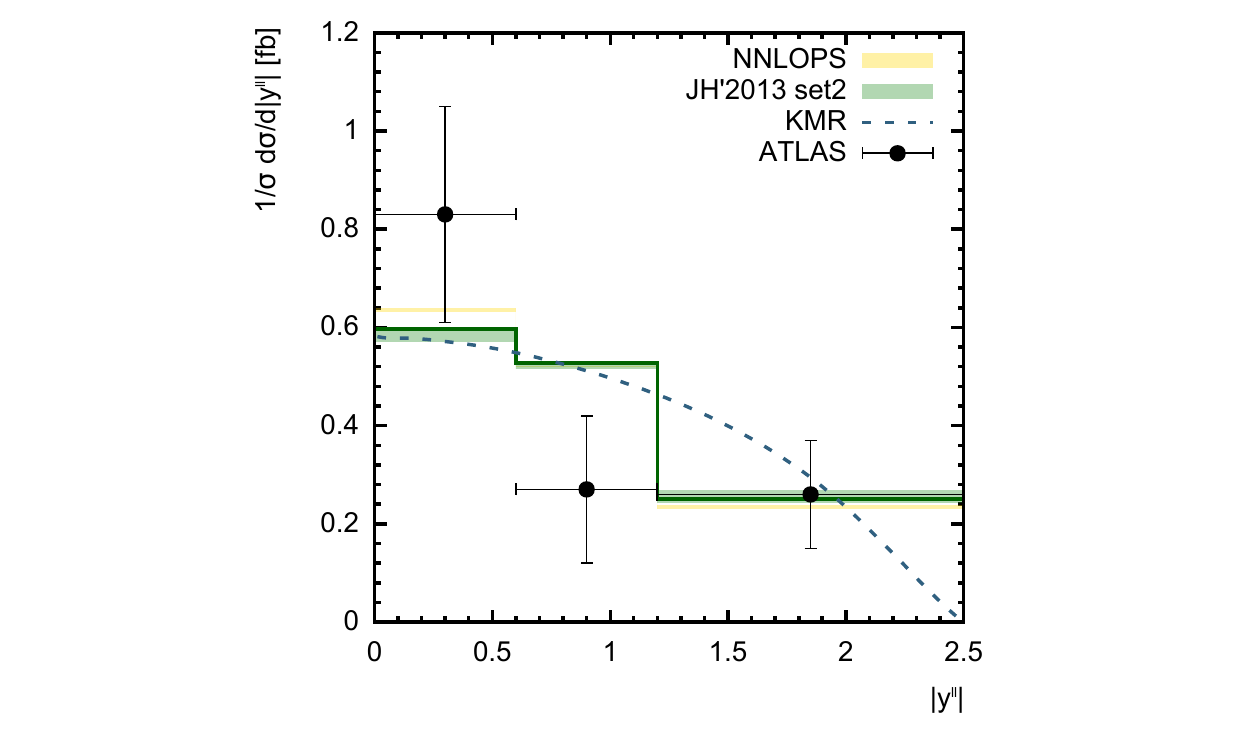}
\caption{The normalized differential cross sections of inclusive Higgs  
production (in the $H \to W^+W^- \to e^\pm \mu^\mp \nu \bar \nu$ decay mode) at $\sqrt s = 8$~TeV
as functions of Higgs transverse momentum and lepton pair rapidity. 
Notation of histograms and curves is the same as in Fig.~1. 
The experimental data are from ATLAS\cite{16}.
The \textsc{nnlops}\cite{34,35} predictions are taken from\cite{16}.}
\label{fig7}
\end{center}
\end{figure}

\begin{figure}
\begin{center}
\includegraphics[width=8cm]{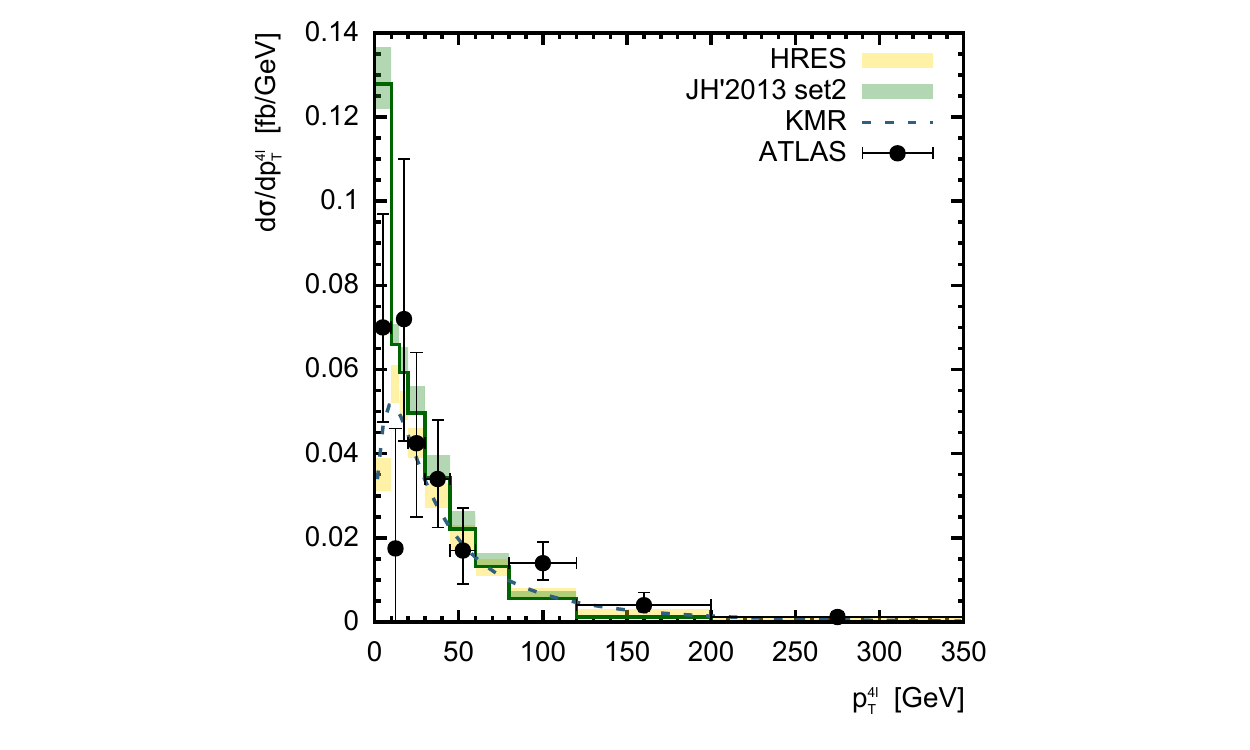}
\includegraphics[width=8cm]{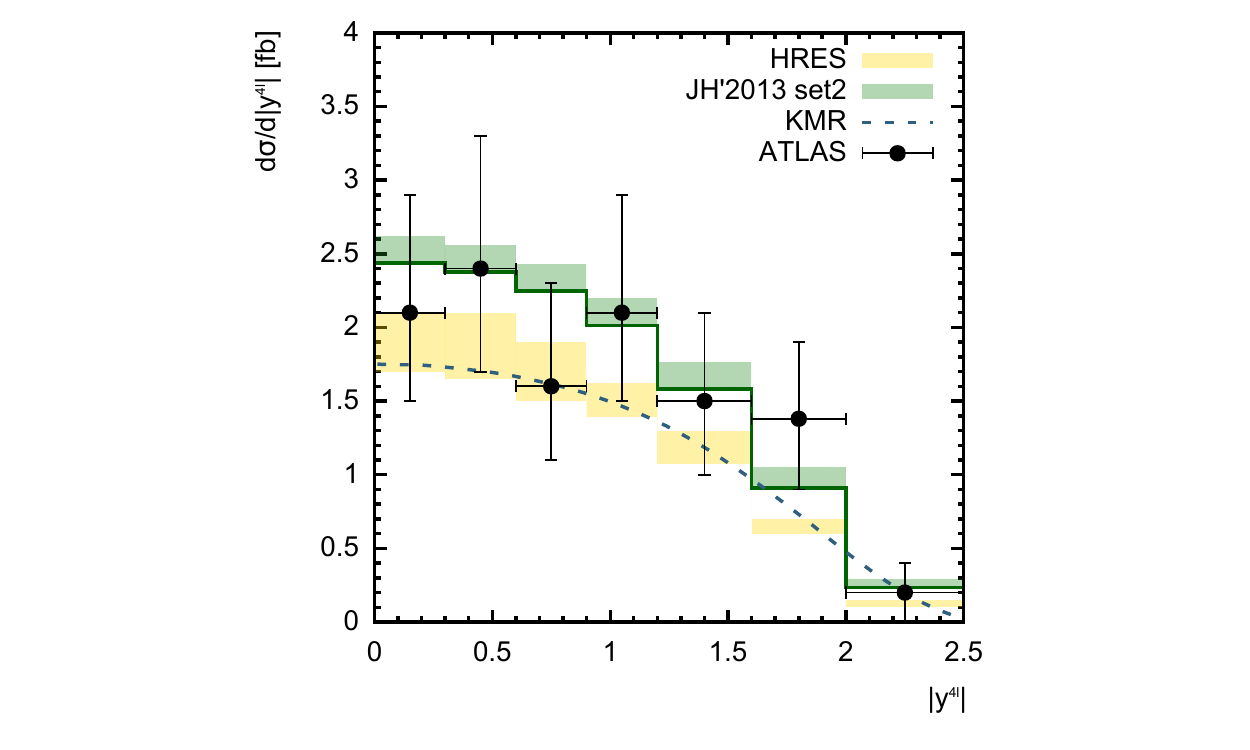}
\includegraphics[width=8cm]{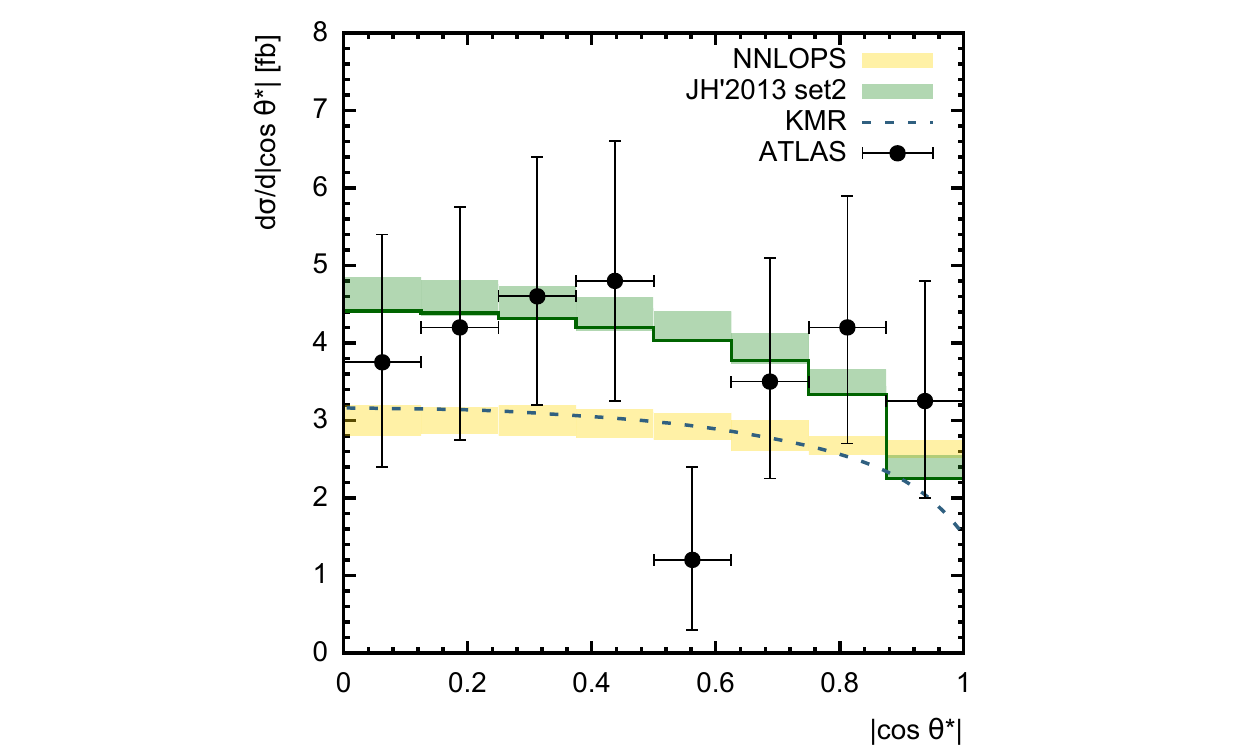}
\includegraphics[width=8cm]{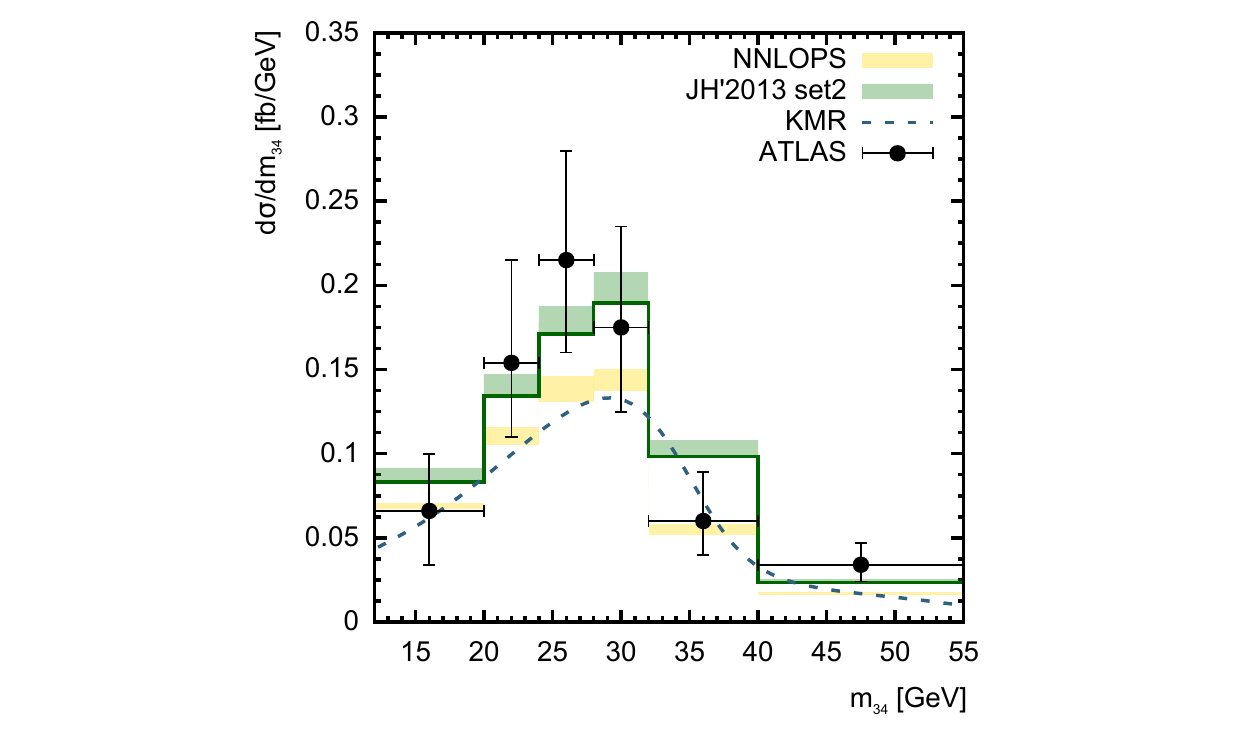}
\includegraphics[width=8cm]{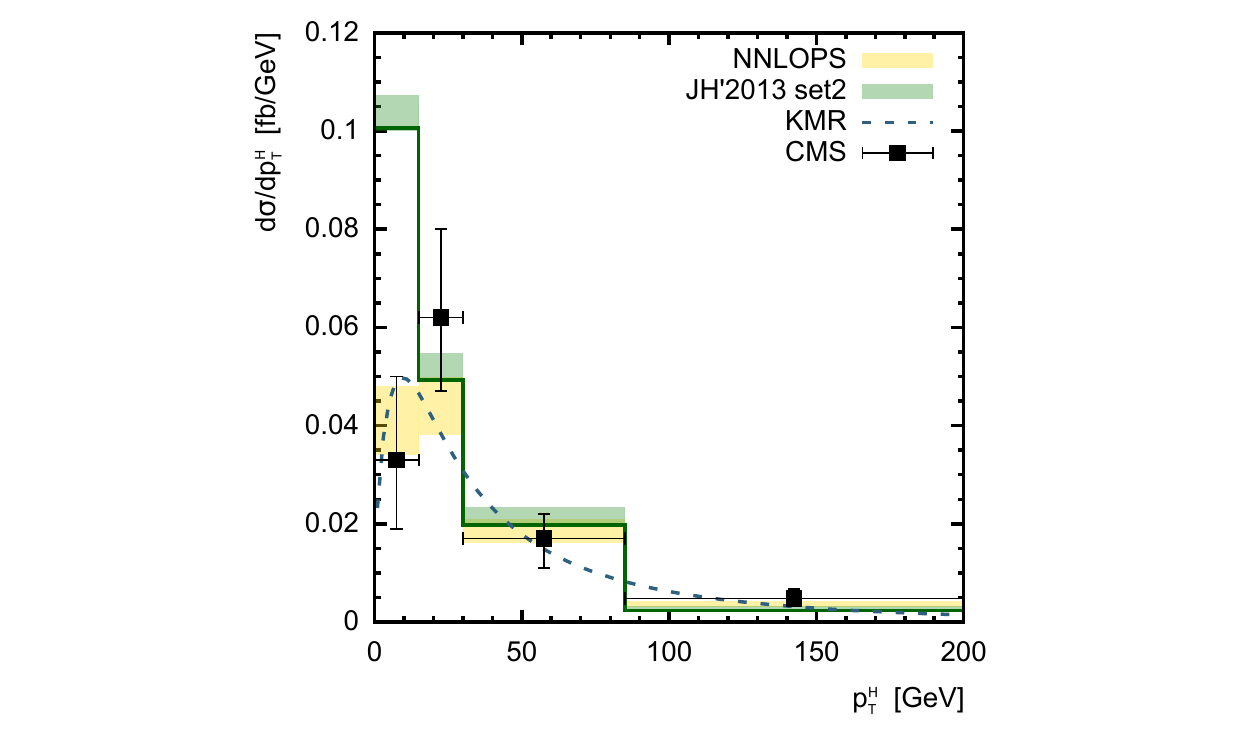}
\caption{The differential cross sections of inclusive Higgs 
production (in the $H \to ZZ^* \to 4l$ decay mode) at $\sqrt s = 13$~TeV
as functions of Higgs boson transverse momentum $p_T^{H}$, rapidity $|y^{H}|$, 
leading lepton pair decay angle $\cos \theta^*$ (in the Collins-Soper frame) and invariant mass $m_{34}$
of the subleading lepton pair. 
Notation of histograms and curves is the same as in Fig.~1. 
The preliminary experimental data are from CMS\cite{18} and ATLAS\cite{20}.
The \textsc{hres}\cite{69} and \textsc{nnlops}\cite{34,35} predictions are taken from\cite{18,20}.}
\label{fig8}
\end{center}
\end{figure}

\end{document}